\documentclass[]{aastex631}
\usepackage[whole]{bxcjkjatype}
\usepackage{comment}

\newcommand{\mjy}{$\mathrm{mJy}\,\mathrm{beam^{-1}}$}
\newcommand{\jy}{$\mathrm{Jy}\,\mathrm{beam^{-1}}$}
\newcommand{\kms}{$\mathrm{km}\,\mathrm{s^{-1}}$}

\newcommand{\msun}{$M_{\odot}$}
\newcommand{\lsun}{$L_{\odot}$}

\newcommand{\two}{I\hspace{-.1em}I}

\shorttitle{Secondary outflow in Ser-emb\,15}
\shortauthors{Sato et al.}
\graphicspath{{./}{figures/}}

\begin{document}

\title{Secondary outflow driven by the protostar Ser-emb 15 in Serpens}

\author[0000-0001-5817-6250]{Asako Sato}
\email{sato.asako.322@s.kyushu-u.ac.jp}
\affiliation{Department of Earth and Planetary Sciences, Graduate School of Science, Kyushu University, 744 Motooka, Nishi-ku, Fukuoka 819-0395, Japan}

\author[0000-0002-2062-1600]{Kazuki Tokuda}
\affiliation{Department of Earth and Planetary Sciences, Faculty of Science, Kyushu University, 744 Motooka, Nishi-ku, Fukuoka 819-0395, Japan}
\affiliation{National Astronomical Observatory of Japan, National Institutes of Natural Sciences, 2-21-1 Osawa, Mitaka, Tokyo 181-8588, Japan}

\author[0000-0002-0963-0872]{Masahiro N. Machida}
\affiliation{Department of Earth and Planetary Sciences, Faculty of Science, Kyushu University, 744 Motooka, Nishi-ku, Fukuoka 819-0395, Japan}

\author[0000-0002-1411-5410]{Kengo Tachihara}
\affiliation{Department of Physics, Nagoya University, Furo-cho, Chikusa-ku, Nagoya 464-8601, Japan}

\author[0000-0002-8217-7509]{Naoto Harada}
\affiliation{Department of Earth and Planetary Sciences, Graduate School of Science, Kyushu University, 744 Motooka, Nishi-ku, Fukuoka 819-0395, Japan}

\author{Hayao Yamasaki}
\affiliation{Department of Earth and Planetary Sciences, Graduate School of Science, Kyushu University, 744 Motooka, Nishi-ku, Fukuoka 819-0395, Japan}

\author[0000-0002-4317-767X]{Shingo Hirano}
\affiliation{Department of Astronomy, School of Science, University of Tokyo, Tokyo 113-0033, Japan}

\author[0000-0001-7826-3837]{Toshikazu Onishi}
\affiliation{Department of Physics, Graduate School of Science, Osaka Metropolitan University, 1-1 Gakuen-cho, Naka-ku, Sakai, Osaka 599-8531, Japan}

\author[0000-0002-9091-963X]{Yuko Matsushita}
\affiliation{National Astronomical Observatory of Japan, National Institutes of Natural Sciences, 2-21-1 Osawa, Mitaka, Tokyo 181-8588, Japan}

\begin{abstract}
We present the detection of a secondary outflow associated with a Class I source, Ser-emb\,15, in the Serpens Molecular Cloud. We reveal two pairs of molecular outflows consisting of three lobes, namely primary and secondary outflows, using ALMA $\mathrm{^{12}CO}$ and SiO line observations at a resolution of $\sim$318\,au. The secondary outflow is elongated approximately perpendicular to the axis of the primary outflow in the plane of the sky. We also identify two compact structures, Sources\,A and B, within an extended structure associated with Ser-emb\,15 in the 1.3\,mm continuum emission at a resolution of $\sim$40\,au. The projected sizes of Sources\,A and B are 137\,au and 60\,au, respectively. Assuming a dust temperature of 20\,K, we estimate the dust mass to be $2.4 \times 10^{-3} \,M_{\odot}$ for Source\,A and $3.3 \times 10^{-4}\,M_{\odot}$ for Source\,B. $\mathrm{C^{18}O}$ line data imply the existence of rotational motion around the extended structure, however, cannot resolve rotational motion in Source\,A and/or B, due to insufficient angular and frequency resolutions. Therefore, we cannot conclude whether Ser-emb\,15 is a single or binary system. Thus, either Source\,A or B could drive the secondary outflow. We discuss two scenarios to explain the driving mechanism of the primary and secondary outflows: the Ser-emb\,15 system is (1) a binary system composed of Source\,A and B or (2) a single star system composed of only Source\,A. In either case, the system could be a suitable target for investigating the disk and/or binary formation processes in complicated environments. Detecting these outflows should contribute to understanding complex star-forming environments, which may be common in the star-formation processes.

\end{abstract}

%% Keywords should appear after the \end{abstract} command. 
%% The AAS Journals now uses Unified Astronomy Thesaurus concepts:
%% https://astrothesaurus.org
%% You will be asked to selected these concepts during the submission process
%% but this old "keyword" functionality is maintained in case authors want
%% to include these concepts in their preprints.

%% From the front matter, we move on to the body of the paper.
%% Sections are demarcated by \section and \subsection, respectively.
%% Observe the use of the LaTeX \label
%% command after the \subsection to give a symbolic KEY to the
%% subsection for cross-referencing in a \ref command.
%% You can use LaTeX's \ref and \label commands to keep track of
%% cross-references to sections, equations, tables, and figures.
%% That way, if you change the order of any elements, LaTeX will
%% automatically renumber them.
%%
%% We recommend that authors also use the natbib \citep
%% and \citet commands to identify citations. The citations are
%% tied to the reference list via symbolic KEYs. The KEY corresponds
%% to the KEY in the \bibitem in the reference list below. 

\section{Introduction} \label{sec:intro}
Protostellar outflows have been detected in early observations around young stellar objects at various evolutionary stages from Class\,0 to Class\,II \citep{EdwardsandSnell1982, Lada1985, Fukui1993, Bachiller1996ARA&A..34..111B, Arce2007prpl.conf..245A}. 
Recent observations have also reported outflow driven from first-core candidates \citep{Chen2010, Fujishiro2020, Maureira2020MNRAS.499.4394M}, brown dwarfs \citep{Whelan2005Natur.435..652W, Phan-Bao2008ApJ...689L.141P, Riaz2019}, and early B-type stars \citep{BeutherandShepherd2005ASSL..324..105B}.
These observations indicate that various types of objects can drive outflows. Early observations also suggest that outflows are ubiquitous phenomena, e.g., in the Milky Way (e.g., \citealt{Wu2004A&A...426..503W, Yang2018ApJS}).
A protostellar (or molecular) outflow can be detected by molecule line emission, such as the rotational transition line of $^{12}$CO. 
The spatial extent of protostellar outflows is much larger than the size of their driving sources (or young stellar objects). 
Thus, detecting molecular outflows is a landmark for discovering young stellar objects in star-forming regions.
In addition, the outflow physical parameters, such as the mass ejection rate and linear and angular momenta, can characterize the physical state of the outflow driving sources (e.g., \citealt{Beuther2002, Matsushita2018}). 
\cite{Wu2004A&A...426..503W} showed a clear correlation between bolometric luminosity and various outflow parameters, such as outflow mass, mechanical luminosity, and force, by using the archival $\mathrm{^{12}CO}$ data where molecular outflows were identified across low- and high-mass sources.
More recent studies presented similar correlations in the Perseus, Ophiuchus, Taurus, Chamaeleon, and Serpens \citep{Yildiz2015}, Orion \citep{Dutta2023arXiv}, and Cygnus-X \citep{Skretas2022} star-forming regions.
Note that the bolometric luminosity is usually used as an index for representing the protostellar mass.
The correlation between bolometric luminosity and outflow parameters has recently been confirmed in the Magellanic Clouds \citep{Tokuda2019ApJ...886...15T, Tokuda2022}.
Therefore, protostellar outflows are significant phenomena for understanding star formation processes across the universe.

In addition to the outflow physical parameters, the outflow morphology is crucial for understanding the outflow driving and star formation processes. 
In a classical picture of star formation, a bipolar conical outflow is expected to be driven by the disk around a protostar \citep[e.g.,][]{1986ApJ...301..571P}. However, recent high-resolution observations have revealed the existence of complex protostellar envelopes that cannot be fully explained by the traditional models of spherical or axisymmetric gas contractions and outflows \citep[e.g.,][]{Tokuda2014ApJ...789L...4T, Tokuda2018ApJ...862....8T, Pineda2020NatAs...4.1158P}. 
While some of these complex systems have been observed to have outflows and internal envelopes with complex shapes due to the effect of the viewing angle \citep{Fernandez2020AJ....159..171F, harada2023ApJ...945...63H}, there are also cases where the outflows have unique morphologies such as mono-poles \citep{Louvet2018, habel2021, sato2023, Hsieh2023ApJ}, quadrupole-like structures \citep{Mizuno1990, Hirano1998, Chen2008ApJ...686L.107C}, or misalignments \citep{Lee2016ApJ, Aso2018, zapata2018, Chuang2021, Okoda2021, hara2021, kido2023ApJ...953..190K}.

Recent high-resolution observations have further revealed a unique morphology of outflows. 
For the first time, \cite{Okoda2021} discovered an additional outflow component surrounding a single protostar instead of a binary system, IRAS 15398-3359: a single-star system drives two pairs of molecular outflows.
They confirmed that the elongation axis for one outflow pair (blue- and red-shifted lobes) is misaligned with (or perpendicular to) the rotation axis of the envelope surrounding the central source, while the elongation direction for the other pair is parallel to the rotation axis of the envelope. 
In summary, they found a quadrupole-like outflow around the single source, in which the outflow axis for one pair of outflows is perpendicular to that of the other pair.
They named the outflow with a propagation direction misaligned with the rotation axis of the envelope or the primary outflow around a single star, the secondary outflow.

Recent magneto hydrodynamic (MHD) simulations have reproduced a misaligned outflow driven by a single protostar (e.g., \citealt{machida2020a, hirano2020}). 
When the rotation axis of the prestellar cloud core is inclined from the global magnetic field, the disk rotation axis or disk normal direction changes with time due to efficient magnetic braking. 
These studies showed that a single protostar could episodically drive outflows, and the direction of the outflows is different in every episode. 
As a result, misaligned or quadrupole-like outflows could be realized around a single protostar (e.g., Fig.~10 of \citealt{machida2020a}). Hence, it is significant to observationally understand the driving mechanism of twin outflows in a single-star system.

In addition to the misaligned outflows in a single-star system, recent observations have shown that outflows driven by a binary system are misaligned with each protostar driving an outflow (e.g., \citealt{zapata2018, hara2021}). 
Numerical simulations have also reproduced twin outflows in a binary system \citep{2017MNRAS.470.1626K,2019MNRAS.486.3647K,saiki2020}. 
Thus, understanding misaligned outflow is also crucial for understanding the formation of binary systems. 
Although the number of observation samples of misaligned outflows has gradually been increasing, more samples are required to understand the diversity of outflows. Protostars grow by mass accretion, and the mass accretion process is closely related to the outflow behavior, which efficiently removes excess angular momentum from the star-forming core, promoting protostellar mass growth and suppressing binary formation \citep[e.g.,][]{2009ApJ...704L..10M}. 
Therefore, we need more observational samples of unusual outflows to reveal the star and binary formation processes. 

Ser-emb\,15 is located within Cluster\,B in the Serpens Molecular Cloud \citep{Harvey2006} at a distance of 436 $\pm$ 9\,pc \citep{Ortiz-Leon2018}.
The Serpens Molecular Cloud is a star-forming region that harbors 34 protostars (9 Class\,0 and 25 Class\,I sources identified by \citealt{Enoch2009}), whereas Ser-emb\,15 was identified as a Class\,I source.
The bolometric temperature, bolometric luminosity, and envelope mass of Ser-emb\,15 were estimated to be 100\,K, 0.4\,\lsun, and 1.3\,\msun\,, respectively \citep{Enoch2011}.
Previous Atacama Large Millimeter/submillimeter Array (ALMA) observations with an angular resolution of $\sim 0 \farcs3$ ($\sim 130$\,au) showed that a single-peaked source traced in the 1.3\,mm continuum emission is associated with Ser-emb\,15 \citep{Francis2019}.
\cite{Bergner2020} presented an image of the $\mathrm{^{12}CO}$\,($J$ = 2--1) line emission toward Ser-emb\,15, using another ALMA data, with an angular resolution of $\sim 0\farcs5$ ($\sim 218$\,au). 
They showed two CO lobes associated with Ser-emb\,15, corresponding to the primary outflow named in this paper. 
Their identified outflow (the primary outflow) is elongated parallel to the rotation axis of the envelope that they also detected in $\mathrm{C^{18}O}$\,($J$ = 2--1), $\mathrm{DCN}$\,($J$ = 3--2) and $\mathrm{H^{13}CN}$\,($J$ = 3--2) line emission.
In \cite{Bergner2020}, we can visually confirm another CO emission elongated toward the eastern direction of Ser-emb\,15, referred to as the secondary outflow in our work.
However, \cite{Bergner2020} did not provide any scientific interpretation of the component, probably because their motivation is not to identify outflows.
In this study, we report the identification of an additional outflow (i.e., secondary outflow) associated with Ser-emb\,15, which should have been mentioned in the previous studies.
We describe the observation details in \S\ref{sec:obs} and our results in \S\ref{sec:results}. 
We discuss the origin of the secondary outflow in \S\ref{sec:dis}. 
Our conclusion is presented in \S\ref{sec:conclusion}.

\section{Observational Imaging} \label{sec:obs}
We used the ALMA archival data for Ser-emb 15 taken from part of the project 2019.1.01792.S (PI: Mardones Diego), including 1.3\,mm continuum, $\mathrm{^{12}CO}$\,($J$ = 2--1; 230.53797\,GHz), SiO\,($J$ = 5--4; 217.10498\,GHz) and $\mathrm{C^{18}O}$\,($J$ = 2--1; 219.560358\,GHz) line data in Band\,6 observed on 2019 November 8, 2020 January 9, and 2021 July 6. 
The observation parameters are summarized in Table\,\ref{para-1}.

The data were reprocessed with the Common Astronomy Software Application (CASA) package \citep{CASA2022PASP..134k4501C} version 6.4.4 and 6.5.0 in the imaging process. We used the \texttt{tclean} task with the hogbom deconvolver for both the continuum and line data. We applied uniform weighting to the continuum data and the Briggs weighting with a robust parameter of 0.5 to the line data.
We continued the deconvolution process on all data until the intensity of the residual image reached the $2\sigma$ noise level. Imaging parameters for the final images, such as synthesized beam size and noise level, are summarized in Table\,\ref{para-2}.

To resolve substructures around the protostar, this study uses the continuum data (Table\,\ref{para-2}) with the highest angular resolution among existing ALMA observations with continuum emission toward Ser-emb\,15.
In addition, the line data in this study has the largest mapping area among existing ALMA observations with CO emission.
We show images of the continuum, CO, SiO, and $\mathrm{C^{18}O}$ line emission from these data in \S\,\ref{sec:results}, \S\,\ref{sec:dis}, and Appendix.

\begin{table}
	\centering
	\caption{Observation parameters}
	\label{para-1}
	\small
	\begin{tabular}{l|lll} % four columns, alignment for each
		\hline
		Project code & \multicolumn{3}{c}{2019.1.01792.S} \\
		\hline
		\hline
		Array & \multicolumn{3}{c}{ALMA 12-m array} \\
		Observation date (YYYY-MM-DD) & 2019-11-08 & 2020-01-09 & 2021-07-06 \\
		Number of antennas & 45& 47& 45\\
		Mapping center (ICRS) & \multicolumn{3}{c}{$18^{h}29^{m}54.3^{s}$, $00$\arcdeg$36'00''.8$}\\
		Field of view (arcsec) & 25.85& 25.85& 24.99 \\
		Mean PWV (mm)& 2.00 & 1.20 & 0.52 \\
		Maximum recoverable size (arcsec) & 9.79& 9.79& 2.45 \\
		Flux calibrator & J2056-4714& J1924-2914& J1924-2914 \\
		Bandpass calibrator & J2056-4714& J1924-2914& J1924-2914 \\
		Phase calibrator & J1851+0035& J1851+0035& J1851+0035 \\
		\hline
	\end{tabular}
\end{table}

\begin{table}
	\centering
	\caption{Summary of imaging parameters}
	\label{para-2}
	\small
	\begin{tabular}{lcccccr} % four columns, alignment for each
		Data set & Weight& \multicolumn{1}{c}{Synthesized beam, Position angle}& Noise level & Velocity resolution \\
		& &[arcsec $\times$ arcsec, deg.] & [\mjy] & [\kms] & Figure\\
		\hline
		continuum & uniform &$0.13 \times 0.091$, $+$84 & 0.24& - & \ref{fig:cont}\\
		$\mathrm{^{12}CO}$\,($J$ = 2--1) & Briggs & $0.96 \times 0.73$, $-$71 & 6.5 & 1.0 & \ref{result-1}a, \ref{fig:outflow}, \ref{fig:shock}, \ref{12co_ch_map_1}, \ref{12co_ch_map_2}\\
 		$\mathrm{^{12}CO}$\,($J$ = 2--1) & Briggs & $0.96 \times 0.73$, $-$71 & 16.0 & 0.1 & \ref{result-1}b, \ref{result-1}c\\
		SiO\,($J$ = 5--4) & Briggs & $1.02 \times 0.82$, $-$78 & 7.0 & 1.0 & \ref{fig:outflow}, \ref{fig:shock}, \ref{sio_ch_blue}, \ref{sio_ch_vsys_red}, \ref{sio_line_spectrum}\\
    $\mathrm{C^{18}O}$\,($J$ = 2--1) & Briggs & $0.99 \times 0.78$, $-$75 & 6.0 & 1.0 & \ref{c18o_ch}\,top\\
    $\mathrm{C^{18}O}$\,($J$ = 2--1) & Briggs & $1.00 \times 0.78$, $-$74 & 30.0 & 0.1 & \ref{c18o_ch}\,bottom\\
 
		\hline
	\end{tabular}
\end{table}

\section{Results} \label{sec:results}
\begin{figure}[ht]
 \centering
  \includegraphics[width=\textwidth]{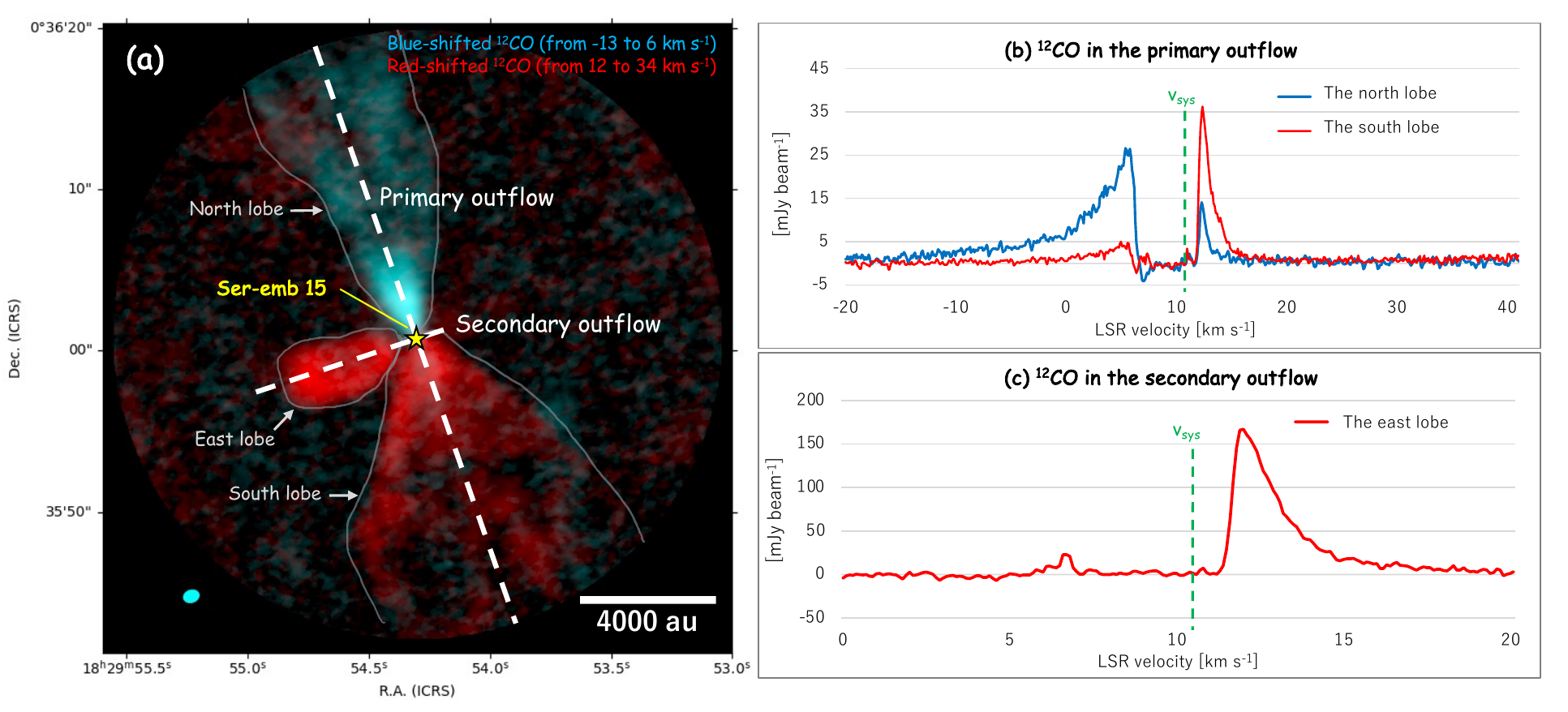}
  \caption{\textbf{(a)} Primary and secondary outflow image. The blue and red components correspond to integrated intensities of the $\mathrm{^{12}CO}$\,($J$ = 2--1) emission larger than $2\sigma$ ($1\sigma = 6.5$\,\mjy) with the LSR velocity in the range from $-$13 to 6\,\kms\,for the blue-shifted component and from 12 to 34\,\kms\,for the red-shifted component. The yellow star symbol represents the location of Ser-emb\,15. The white dashed lines represent the axis of the primary outflow with P.A. = $+$20 degrees, along which the blue- and red-shifted CO emission is elongated from north to south, and the axis of the secondary outflow with P.A. = $+$110 degrees, along which the red-shifted CO emission is elongated to the east toward the center. The light blue ellipse at the bottom-left corner of the panel represents the synthesized beam size of the $\mathrm{^{12}CO}$\,($J$ = 2--1) emission.
  \textbf{(b)} $\mathrm{^{12}CO}$\,($J$ = 2--1) line spectra in the primary outflow. The solid blue and red lines correspond to the spectrum of the emission integrated over the north and south lobes, respectively, as indicated in panel (a), with respect to Ser-emb\,15. 
  \textbf{(c)} $\mathrm{^{12}CO}$\,($J$ = 2--1) line spectrum in the secondary outflow. The solid red line corresponds to the spectrum of the emission integrated over the east lobe, as indicated in panel (a), with respect to Ser-emb\,15 . 
  The dashed green line in panels (b) and (c) corresponds to the systemic velocity of 10.5\,\kms.
  The systemic velocity is estimated from the $\mathrm{C^{18}O}$\,($J$ = 2--1) line spectrum presented in Figure\,\ref{c18o_ch} of Appendix\,\ref{app_C18O}.
  }
  \label{result-1}
\end{figure}

Figure\,result-1 shows a reference of the primary and secondary outflows, presenting blue- and red-shifted $\mathrm{^{12}CO}$\,($J$ = 2--1) emission and $\mathrm{^{12}CO}$\,($J$ = 2--1) line spectra \footnote{To derive line spectra presented in Figures~\ref{result-1}, \ref{c18o_ch}, and \ref{sio_line_spectrum}, we utilized an image visualization and analysis tool the Cube Analysis and Rendering Tool for Astronomy (CARTA, \citealt{CARTA2021zndo...3377984C}) version 4.0.0-beta.1.}.
The central yellow star symbol in Figure\,\ref{result-1}a represents the infrared source Ser-emb\,15 identified in the Cores to Disks (c2d) Spitzer Legacy Program \citep{Evans2003}.
With respect to Ser-emb\,15, the CO emission is extended in the north, south, and east directions.
This paper calls the pair of northern and southern CO components `primary outflow' and the eastern CO component `secondary outflow.'
Figure\,\ref{result-1}b shows that the northern and southern components have blue- and red-shifted wing-like lobes, respectively, indicating that the CO emission originates from the (primary) outflow. 
Figure\,\ref{result-1}c shows that the eastern component has a red-shifted wing-like lobe, indicating that the CO emission originates from the (secondary) outflow. 
However, we could not detect the opposite side lobe of the secondary outflow, which is expected to extend to the western side of Ser-emb\,15. Possibilities of the non detection are discussed in Section~\ref{sec:non-detect}.

In \S\ref{result-cont}, we present a 1.3\,mm continuum image obtained from the ALMA 12-m array.
In the image, we identify two dense sources and estimate the physical properties of the sources, such as projected sizes. 
In \S\ref{result-outflow}, we present $\mathrm{^{12}CO}$\,($J$ = 2--1) and SiO\,($J$ = 5--4) images obtained from the ALMA 12-m array. 
In the CO images, we identify molecular outflows and derive their physical properties, such as lengths and velocity ranges. 
We also identify the high-velocity flow (or jets) and shocked gas traced by the SiO\,($J$ = 5--4) line emission and estimate their physical properties.

\subsection{1.3 mm continuum} \label{result-cont}

\begin{figure}[ht]
  \centering
  \includegraphics[width=\textwidth]{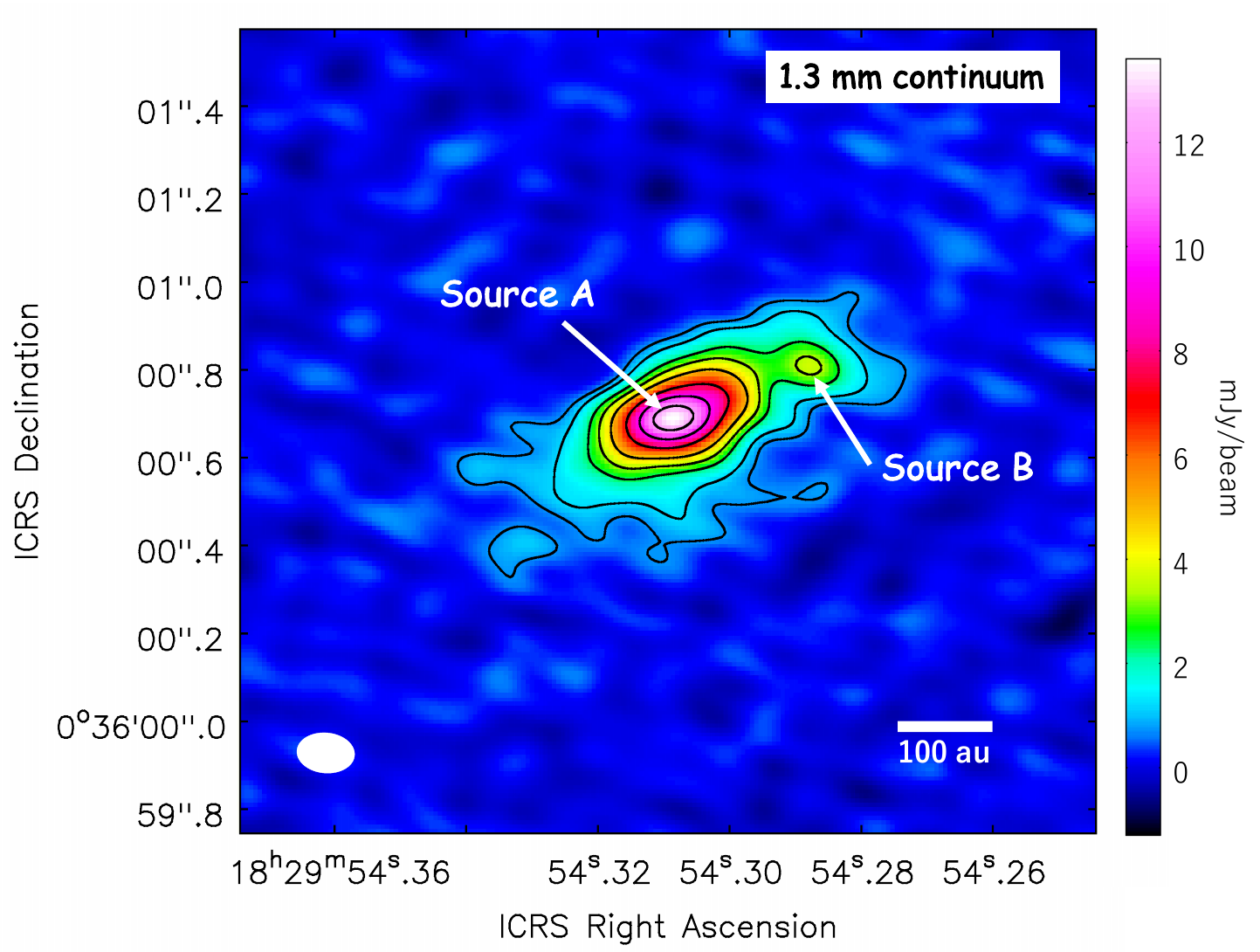}
  \caption{ALMA 12-m array 233\,GHz continuum image represented by color scale and black contours. The black contour levels are [3, 5, 10, 14, 20, 30, 40, 50] $\times 1\sigma$ ($1\sigma = 0.24$\,\mjy). The white ellipse at the bottom-left corner is the largest synthesized beam size of the continuum emission.}
  \label{fig:cont}
\end{figure}

Figure~\ref{fig:cont} presents a 1.3\,mm continuum image obtained from the ALMA 12-m array with, at least, five times smaller beam area compared with that of earlier mm/sub-mm observations \citep{Enoch2007, Enoch2011, Francis2019, Bergner2019}.
We detected an extended structure elongated in the north-west to south-east direction, whose spatial distribution is consistent with millimeter continuum images taken with the CSO/Bolocam \citep{Enoch2007}, CARMA \citep{Enoch2011}, and ALMA \citep{Francis2019, Bergner2019}.
The structure has two peaks within the $3\sigma$ contour. 
We identified two continuum sources from 2D Gaussian fitting using a CASA task of \texttt{imfit}, and named the east source `Source\,A' and the west source `Source\,B', as shown in Figure~\ref{fig:cont}. 
Source\,B was identified for the first time in this study. 
The peak position, peak flux, projected size, position angle (hereafter P.A.), and integrated flux measured by 2D Gaussian fitting for each source are summarized in Table\,\ref{cont-fit}.
The absolute position accuracy of the peak position was derived using the recommended calculation scheme, which is the formula of beam$_\mathrm{(FWHM)}$/S/N/0.9 (see the ALMA technical handbook).

\begin{table}[]
  \centering
  \footnotesize
  \caption{Physical properties for Source\,A and Source\,B obtained from the 1.3\,mm continuum image. The position of Source\,A and B are also confirmed in Figure\,\ref{fig:cont}.}
  \label{cont-fit}
  \begin{tabular}{c|cccccc}
  Name & Peak position & Position accuracy & Peak flux & Projected size & Position angle & Integrated flux \\
  \, & (ICRS R.A., Dec.) & [arcsec] & [\mjy] & (FWHM) [arcsec] & [degree] & [mJy] \\
   \hline
  Source\,A& $18^{h}29^{m}54.307^{s}$, $-00$\arcdeg$36'00''.696$ & 0.022 & 11.23 &0.316$\times$0.129 & $+$116.9 & 52.4 \\
  Source\,B& $18^{h}29^{m}54.288^{s}$, $-00$\arcdeg$36'00''.812$ & 0.074 & 3.594 & 0.139$\times$0.088 & $+$78.2 & 7.28 
  \end{tabular}
\end{table}

Assuming that the 1.3\,mm continuum emission comes from optically thin dust thermal emission and that the temperature distribution of the continuum source is uniform, we can estimate the dust mass as
\begin{equation}
    M_{\mathrm{dust}} = \frac{F_{\mathrm{\lambda}}d^2}{\kappa_{\mathrm{\lambda}}B_{\mathrm{\lambda}}(T_{\mathrm{dust}})},
\end{equation}
where $\kappa_{\mathrm{\lambda}} = 0.899\,\mathrm{cm^2\,g^{-1}}$ \citep{ossenkopf1994} is the dust opacity, $B_{\mathrm{\lambda}} $ is the Planck function for the dust temperature $T_{\mathrm{dust}} = 20\,\mathrm{K}$ \citep{Andrews2005ApJ}, $F_{\mathrm{\lambda}}$ is the total flux density for the continuum source emission, and $d$ = 436\,pc \citep{Ortiz-Leon2018} is the distance to the source.
The dust mass for Source\,A is estimated to be $M_{\rm dust, A} \simeq 2.4 \times 10^{-3} \,M_{\odot}$ and the one for Source\,B is estimated to be $M_{\rm dust, B} \simeq 3.3 \times 10^{-4}\,M_{\odot}$.

We also estimated total $\mathrm{H_2}$ column density for Sources\,A and B using the above dust masses with a molecular weight of 2.37 and an assumption of the dust-to-gas mass ratio of 1:100, referring to \cite{Bergner2020}.
The $\mathrm{H_2}$ column density is derived to be $6.3\times10^{25} \mathrm{cm^{-2}}$ for Source\,A and $2.9\times10^{25} \mathrm{cm^{-2}}$ for Source\,B. \cite{Bergner2019} also estimated the total $\mathrm{H_2}$ column density for Ser-emb\,15 to be $1.3\times10^{24} \mathrm{cm^{-2}}$. 
The difference between them could be caused by different settings and observational results such as mass sensitivity, frequency, dust opacity and estimated source size.
In particular, the source size is considered to be the primary cause of the difference in the column density.
\cite{Bergner2020} estimated the source size as a single component for Ser-emb\,15 using 2D Gaussian fitting. 
On the other hand, this study performed the same fitting for two spatially resolved components, i.e., Sources\,A and B, both of which are located within the component Ser-emb\,15 identified in \cite{Bergner2020}. 
The measured areas for Sources\,A and B ($\sim$0\farcs3$\times$0\farcs1) are approximately 20 times smaller than the area for Ser-emb\,15 ($\sim$1\farcs2$\times$0\farcs5; \citealt{Bergner2020}). 
Therefore, the difference in the source sizes should result in an order of magnitude difference in the column densities.
In addition, we further discuss how uncertainties, such as dust opacity and temperature, influence the estimated dust mass in Appendix\,\ref{app_dust}.

\subsection{Molecular lines} \label{result-outflow}

\begin{figure}[ht]
  \centering
  \includegraphics[width=\textwidth]{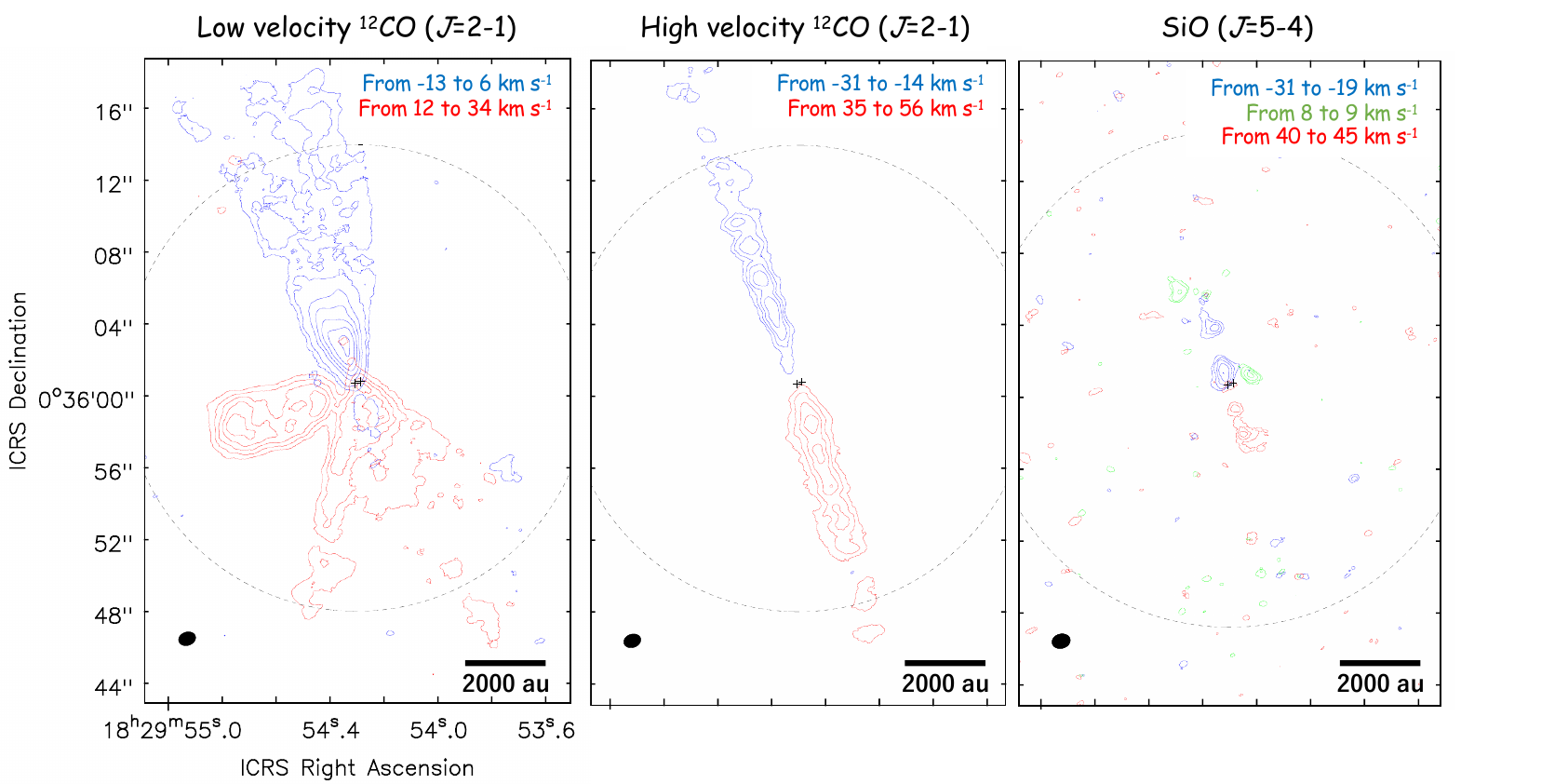}
  \caption{$^{12}$CO($J$=2--1) and SiO($J$=5--4) integrated intensity maps. The two black crosses in each panel correspond to the peak positions of the dust continuum sources, Sources\,A (east cross) and B (west cross). The primary beam at full width half maximum is also depicted by a dashed black circle in each panel.
  \textbf{Left}: Low-velocity components of $^{12}$CO($J$=2--1) emission. The blue and red contours represent the integrated intensity of the blue- and red-shifted $^{12}$CO($J$=2--1) emission with the LSR velocity in the range from $-$13 to 6\,\kms and from 12 to 34\,\kms, respectively. The blue contour levels are [2, 4, 10, 15, 20, 25, 30]$\times 1\sigma$ ($1\sigma = 60.0$\,\mjy). The red contour levels are [2, 4, 6, 8, 10]$\times 1\sigma$ ($1\sigma = 80.0$\,\mjy). The black ellipse at the bottom-left corner represents the synthesized beam size of the $^{12}$CO($J$=2--1) emission. 
  \textbf{Center}: High-velocity components of $^{12}$CO($J$=2--1) emission. The blue and red contours represent the integrated intensity of the blue- and red-shifted $^{12}$CO($J$=2--1) emission in the LSR velocity ranges from $-$31 to $-$14\,\kms\,and from 35 to 56\,\kms, respectively. The blue contour levels are [2, 4, 6, 8]$\times 1\sigma$ ($1\sigma = 68.9$\,\mjy). The red contour levels are [2, 4, 6, 8]$\times 1\sigma$ ($1\sigma = 84.2$\,\mjy). The black ellipse at the bottom-left corner is the same as that in the left panel of this figure.
  \textbf{Right}: SiO($J$=5--4) emission. The blue and red contours represent the integrated intensity for the blue- and red-shifted SiO($J$=5--4) emission in the LSR velocity ranges from $-$31 to $-$19\,\kms and from 40 to 45\,\kms, respectively. The green contours represent the integrated intensity of the SiO emission around the systemic velocity in the LSR velocity range from 8 to 9\,\kms. The blue contour levels are [2, 4, 6, 8, 10]$\times 1\sigma$ ($1\sigma = 40.0$\,\mjy). The red contour levels are [2, 4, 6, 8, 10]$\times 1\sigma$ ($1\sigma = 20.0$\,\mjy). The green contour levels are [2, 4, 6, 8, 10]$\times 1\sigma$ ($1\sigma = 7.0$\,\mjy). The black ellipse at the bottom-left corner represents the synthesized beam size for the SiO($J$=5--4) emission.}
  \label{fig:outflow}
\end{figure}

Figure\,\ref{fig:outflow} shows the integrated intensity images of the low-velocity $^{12}$CO\,($J$=2--1) (left), high-velocity $^{12}$CO\,($J$=2--1) (middle), and SiO\,($J$=5--4) (right) line emission.
The channel maps of both line emission are presented in Appendix\,\ref{app_CO} and \ref{app_SiO}.
Note that we adopted a systemic velocity for Ser-emb\,15 of 10.5\,\kms based on the line spectrum of $\mathrm{C^{18}O}$\,($J$ = 2--1) data obtained from the same project as the CO and SiO data (see Figure\,\ref{c18o_ch} of Appendix~\ref{app_C18O}).

\subsubsection{$^{12}$CO\,($J$=2--1)} \label{result_co}
The left panel in Figure\,\ref{fig:outflow} shows the low-velocity CO components, which comprise three parts: the blue-shifted CO emission extending to the north-east, the red-shifted CO emission extending to the south-west, and the red-shifted CO emission extending to the south-east.
The north-east component has a length of 18$''$ ($\sim$ 7848\,au) with P.A. = $+$20 degrees and an LSR velocity range from $-$13 to 6\,\kms.
The south-west component has a length of 7.5$''$ ($\sim$ 3217\,au) with P.A. = $+$200 degrees and an LSR velocity range from 12 to 34\,\kms.
The south-east component has a length of 8.5$''$ ($\sim$ 3646\,au) with P.A. = $+$110 degrees and an LSR velocity range from 12 to 23\,\kms.
The blue-shifted CO emission from the north-east component seems to wiggle and is more collimated than the red-shifted counterpart (the red-shifted CO emission from the south-west component).
In addition, the south-east component has an elliptical-shaped structure, in contrast to the corn-shaped structures of the north-east and south-west components. 
Within the elliptical-shaped structure, we can see an intensity peak at the eastern edge corresponding to the tip of the secondary outflow (see also the channel maps in Appendix\,\ref{app_CO}).
These three CO components are associated with Source\,A and B described in \S\ref{result-cont}. 
Note that the black crosses shown in Figure\,\ref{fig:outflow} denote the peak positions of Source\,A and B in the continuum emission.
Hence, we identified these CO components as molecular outflows and referred to the north-east and south-west components as the `primary outflow' and the south-east component as the `secondary outflow.'

The spatial extent of the primary outflow corresponds to the previously identified CO outflows \citep{Bergner2020}. 
The axis of the primary outflow detected in this study is perpendicular to the major axis of the rotating envelope reported by \cite{Bergner2020}.
The secondary outflow is identified as a molecular outflow for the first time in this study.

The central panel of Figure\,\ref{fig:outflow} shows the high-velocity $^{12}$CO\,($J$=2--1) components of the primary outflow.
We detected collimated structures for both blue- and red-shifted CO emission extending from north-east to south-west.
The north-east component has a size of 18$''$ ($\sim$ 7722\,au) with P.A. = $+$20 degrees and an LSR velocity range from $-$31 to $-$14\,\kms.
The south-west component has a size of 14\farcs7 ($\sim$ 6306\,au) with P.A. = $+$200 degrees and an LSR velocity range from 35 to 56\,\kms. 
The position angles for these two components are the same as those for the low-velocity CO components of the primary outflow shown in the left panel of Figure\,\ref{fig:outflow}. 
Thus, these high-velocity flows are considered to be associated with the primary outflow.
The high-velocity components are more collimated than the lower-velocity components. 
In addition, we can see several knots within the high-velocity components. 
These features, such as collimated and knot-like structures, are similar to those for extreme high velocity (EHV) jets (e.g., \citealt{Tychoniec2019, 2019ApJ...871..221M}).
The blue-shifted component has at least four knots with an average separation of 1\farcs7 ($\sim$729\,au).
The red-shifted component also has at least four knots with an average separation of 1\farcs8 ($\sim$772\,au).
Given the average separations, maximum LSR velocities, and outflow inclination angle of $\sim$30\,degree expected from the aspect ratio of the continuum components (Sources\,A and B), we can infer an average duration of the episodic mass ejection event of $\sim$50 years for both the red- and blue-shifted CO components. 
The expected duration of the mass ejection for the primary outflow is comparable to that for the EHV jet in OMC-3/MMS5 ($\sim$50\,years estimated by \citealt{2019ApJ...871..221M}), and also for protostellar jets in the ALMA Survey of Orion PGCCs (ALMASOP) samples ($\sim$20 to 175\,years estimated by \citealt{Dutta2023arXiv}).

\subsubsection{SiO\,($J$=5--4)} \label{result_sio}
%%The SiO emission could trace various shocked regions where the outflows are considered to interact with the surrounding gas. 
The SiO emission has been detected as a tracer of collimated outflows \citep{Mikami1992ApJ...392L..87M, zhang2002, zapata2006, hirano2010, liu2021}, an extended bow shock of the protostellar outflow \citep{gueth1998, shimajiri2008}, and local shocked regions along the outflow cavity wall \citep{Codella2017A&A...605L...3C, sato2023}.
SiO in the gas phase is believed to be formed by the destruction or sputtering of silicate grains \citep{ziurys1989, caselli1997}. 

The right panel of Figure\,\ref{fig:outflow} shows SiO\,($J$=5--4) emission associated with Ser-emb\,15, which is reported for the first time in this study. In the image, both the high-velocity (red and blue) and low-velocity (approximate systemic velocity, green) emission is plotted. The line spectrum of SiO \,($J$=5--4) is presented in Appendix\,\ref{app_SiO}.
The blue-shifted SiO emission with high velocity consists of two components, denoted by the blue contours in the right panel.
One blue-shifted SiO component is directly associated with Sources\,A and B with a size of 1\farcs6 ($\sim$ 686\,au) with P.A. = $+$20 degrees and an LSR velocity range from $-$31 to $-$19\,\kms.
The other blue-shifted SiO component is in the north-east, with its flux peak located 3\farcs3 ($\sim$ 1415\,au) from Sources\,A and B with P.A. = $+$20 degrees and an LSR velocity range from $-$27 to $-$23\,\kms.
The red-shifted SiO emission, denoted by the red contours in the right panel, consists of one component located in the south-west toward the peak for Sources\,A and B. Its projected distance is approximately 1\farcs2 ($\sim$ 514\,au) from Sources\,A and B with a size of 3$''$ ($\sim$ 1287\,au) and P.A. = $+$200 degrees with an LSR velocity range from 40 to 45\,\kms.
The red-shifted SiO component has two distinct intensity peaks in the high-velocity SiO emission. 
The high-velocity red- and blue-shifted components show an approximately point-symmetric distribution with respect to Sources\,A and B.
%%This red-shifted SiO component exhibits two distinct intensity peaks within, demonstrating an approximately point-symmetric distribution relative to Sources\,A and B in the high-velocity SiO emission.

The blue- and red-shifted SiO emission is associated with the high-velocity CO components of the primary outflow because the position and velocity ranges for the SiO emission roughly coincide with those for the CO emission. 
We could not detect SiO components with a similar velocity range to the low-velocity CO components.

In addition, near the systemic velocity of 10.5\,\kms, we detected two extended components in the SiO\,($J$=5--4) emission in the LSR velocity range from 8 to 9\,\kms (the green contours in the right panel).
One component is located 1\farcs4 ($\sim$ 600\,au) west from the crosses with a size of 1\farcs3 ($\sim$ 557\,au).
This component is located on the opposite side of the red-shifted lobe of the secondary outflow with respect to Sources\,A and B (see Fig.\,\ref{result-1} right).
Based on the non-detection of other CO lobe of the secondary outflow at the western side with respect to Ser-emb\,15, this SiO component may trace a shocked region arising from a collision of gas ejected from Ser-emb\,15 with the surrounding material at the western side of Ser-emb\,15 (for detailed discussion see \S\ref{sec:non-detect}).
The other low-velocity SiO component is located 5\farcs8 ($\sim$ 2,488\,au) north-east from the continuum sources with an extension of 1\farcs1 ($\sim$ 471\,au) and is within the northern CO lobe of the primary outflow.
This component is not along the major axis of the primary outflow, and the velocity range of 8 to 9\,\kms\, does not coincide with that of the CO outflow presented in the left and central panels of Figure\,\ref{fig:outflow}. These findings imply that the low-velocity SiO components may trace the shocked regions resulting from the collision between the primary outflow and surrounding material. 

\vspace{3pt}
It should be noted that we used the CO emission data covering the most extensive (or largest) area toward Ser-emb\,15 among the past ALMA observations.
The data indicates that the primary outflow extends beyond the observation area, as seen in Figure\,\ref{fig:outflow}.
Thus, follow-up observations are required covering a broader area to identify the outflows correctly.
In addition, we detected a red-shifted SiO emission at an LSR velocity of $\sim45$\,\kms, which is at the edge of the spectral window, indicating that the SiO emission can be detected at a higher LSR velocity than 45\,\kms.
Thus, follow-up observations are also necessary to appropriately detect SiO emission in this region.

\section{Discussion} \label{sec:dis}
In \S\ref{sec:results}, we presented two pairs of outflows consisting of three lobes, i.e., the primary and secondary outflows, which are considered to be driven from a system, Ser-emb\,15. 
Our continuum data spatially resolved the central clump of the system and detected two sources, Sources\,A and B (see \S\ref{result-cont}).

We also clearly detected a $\mathrm{C^{18}O}$\,($J$ = 2--1) emission associated with the central continuum clump of Ser-emb\,15 (see Figure\,\ref{c18o_ch} in Appendix~\ref{app_C18O}).
The $\mathrm{C^{18}O}$ emission showed a rotational motion around the minor axis of the structure detected in the continuum emission.
The rotational motion or velocity gradient can also be confirmed in the east and west directions across the central continuum sources, Sources\,A and B. 
The rotational motion detected in this study would correspond to that observed in a previous study \citep{Bergner2020}.
The $\mathrm{C^{18}O}$\,($J$ = 2--1) emission data, however, does not have a sufficient angular resolution to spatially resolve the separation between the continuum sources, Sources\,A and B ($\sim$ 0\farcs5).
Thus, we could not determine whether the rotational motion detected in $\mathrm{C^{18}O}$\,($J$ = 2--1) originates from Source\,A and/or Source\,B or the common envelope of the sources.
Therefore, there is still uncertainty as to whether Ser-emb\,15 is a binary or a single system.
In Section\,\ref{sec:binary} and \ref{sec:single}, we discuss the likely origins of the primary and secondary outflows assuming a binary or a single star system. Section\,\ref{sec:non-detect} discusses uncertainties affecting the dust mass estimated in Section\,\ref{result-cont}.

\subsection{Binary system scenario} \label{sec:binary}
If each source (Sources\,A and B) contains a protostar, this system is binary with the separation of 218\,au.
If this is the case, the primary and secondary outflows could be driven by the circumstellar disk (CSD) of each protostar or the circumbinary disk (CBD). As we could not confirm any sign of a protostar embedded in Sources\,A and B, we must consider the origin of the outflows in a binary formation.

\subsubsection{Both outflows driven by Source\,A}
First, we consider a scenario in which both the primary and secondary outflows are driven by Source\,A and the outflows with different directions originate from a directional change of the circumstellar disk around Source A. Outflows should appear in the direction normal to the circumstellar disk \citep{machida2020a}. A directional change of the disk normal could occur when the circumstellar disk around the primary star interacts with that around the companion star. Thus, the direction normal to the disk around Source\,A could change due to the companion Source\,B. When the direction normal to the disk driving the outflows changes, the outflow direction should also change \citep{2013MNRAS.428.1321T}. In this case, the misalignment between the outflow axes for the primary and secondary outflows can be attributed to an interaction between protostars (Sources\,A and B). A directional change of the outflows caused by binary (or multiple) interactions has been reported for Orion KL \citep{bally2015, bally2020} and HW2 \citep{Cunningham2009}. For example, \cite{Cunningham2009} found a precessing jet with four distinct jet axes in HW2 with $\mathrm{H_2}$ 2.12\,$\mu$m line emission. They proposed that the orientation of the jet is changed by $\sim$45\,degrees in $\sim10^4$\,years due to the disk precession caused by the interaction between HW2 and its sibling star.

\subsubsection{Each outflow driven by Sources\,A or B}
Next, we consider that the primary and secondary outflows are driven by Source\,A and B, respectively. 
We confirmed that the total flux of Source\,B is very weak compared with that of Source\,A by seven times (see Table~\ref{cont-fit}). Thus, the protostar and circumstellar disk embedded in Source B are expected to be less massive than those of Source\,A. Additionally, the velocity of the secondary outflow is slower than that of the primary outflow. Given the correlation between bolometric luminosity and outflow energy described in \S\ref{sec:intro} (e.g., \citealt{Wu2004A&A...426..503W}), Source\,B is considered to drive a less energetic outflow than Source\,A, i.e., the secondary outflow. Recent observations have reported misaligned outflows driven from a binary system \citep{zapata2018, hara2021}, in which each protostar drives an outflow. Theoretical studies have suggested that misalignment of outflow axes can be explained by either turbulent fragmentation \citep[e.g.,][]{Bate2000MNRAS.317..773B, 2013MNRAS.428.1321T} or a magnetic field inclined with respect to the disk rotation axis (\citealt{2019MNRAS.486.3647K}, Saiki et al. in prep).

\subsubsection{Origin of the binary formation}
We should also consider the origins of the binary formation in Ser-emb\,15 through theoretically proposed fragmentation processes: disk fragmentation occurring due to gravitational instability in a rotating disk \citep{Adams1989ApJ, Stamatellos2009MNRAS, Kratter2010ApJ}, and turbulent fragmentation occurring within a turbulent core during the pre-stellar phase \citep{Padoan2002ApJ, Offner2010ApJ, Tokuda2014ApJ...789L...4T}. 
Recent observations have suggested that disk fragmentation is plausible for forming close multiple systems with separations of less than $\sim600$ au (e.g., \citealt{Tobin2018ApJ}). 
In contrast, turbulent fragmentation could form wide multiple systems with separations of larger than $\sim$1000\,au (e.g., \citealt{Pineda2015Natur, Lee2016ApJ}). Recent observations with the Submillimeter Array (SMA) have reported that the outflow orientations in binary pairs are randomly distributed with separations from 1000\,au to 10,000\,au \citep{Lee2016ApJ}.
In other words, the binary systems have misaligned outflows, resulting from the complex angular momentum. 
Such outflow orientations in the binary systems could be realized in the turbulent fragmentation case.

The Ser-emb 15 could be a binary system with a separation of $\sim$218 au and exhibits misaligned outflows.
Based on the previous studies, it is difficult to identify the formation mechanism for the Ser-emb 15 system with current existing data.
Disk fragmentation is expected from the perspective of binary separation, while turbulent fragmentation may be plausible from the perspective of the orientation of outflows.
The Ser-emb 15 system is classified as Class I in the SED analysis \citep{Enoch2009}.
Thus, if this system is formed by turbulent fragmentation, the interaction between the binary pairs can shrink the separation into a close binary system in a further evolutionary stage (see Fig.3 of \citealt{2019MNRAS.486.3647K}). 
Moreover, considering the uncertainty in the inclination of the outflows and the orientation of the rotation axes of the disks in Sources\,A and B due to the low angular resolution of the continuum and C$^{18}$O data, it is also difficult to identify the origin of the fragmentation process or the binary formation mechanism from our observational results.
Thus, we need follow-up higher resolution observations toward the same target.

\subsection{Single star system scenario} \label{sec:single}
If only a single protostar is embedded in the system, both the primary and secondary outflows are driven by a single object. This subsection discusses the possibilities to explain the misaligned outflows in a single-star system.

\subsubsection{Outflows driven at a same epoch}
Firstly, we consider that both the primary and secondary outflows can be driven by a protostar embedded in Source\,A at the same epoch. As shown in \S\ref{result-cont}, the continuum emission shows an asymmetric structure with a $3\sigma$ contour in the system, in which the emission from Source\,A is much more dominant than that from Source\,B. Thus, the asymmetry could be attributed to a warped disk, in which the outer disk is inclined relative to the inner disk (e.g., \citealt{Sakai2018, Sai2020}). In other words, the direction normal to the inner disk differs from that for the outer disk within Source\,A. 

Recent MHD simulations have suggested that outflows are driven by a magnetic pressure gradient force in the innermost disk region, because the magnetic field is coupled with the neutral gas and can be amplified after the magnetic field in the disk dissipates \citep[e.g.,][]{Machida2008,2013ApJ...763....6T,2018A&A...615A...5V}. \citet{2002ApJ...575..306T} have shown that a well-collimated outflow configuration is realized when the magnetic pressure gradient force drives the flow. Recent simulations have also shown that the outflow driven from the innermost disk region is more collimated than that driven from the intermediate disk region \citep[see review by][]{2022arXiv220913765T}. In addition, the inner outflow should have a high velocity because the outflow velocity roughly corresponds to the Keplerian velocity at the outflow driving radius \citep{1961MNRAS.122..473M,1968MNRAS.138..359M}. The Keplerian velocity increases as the radius from the protostar decreases. On the other hand, the outer outflow is considered to be driven by a magnetic centrifugal force in the intermediate disk region with a wide opening angle and relatively low velocity (e.g., \citealt{1982MNRAS.199..883B, Bjerkeli2016Natur}). Furthermore, the high-velocity components driven near the protostar can show episodic mass ejection \citep{machida2014ApJ...796L..17M, machida_basu2019ApJ...876..149M}, as we have also shown in the central panel of Figure\,\ref{fig:outflow}. These features of the outflow derived in theoretical and simulation studies agree well with the characteristics of the primary outflow. We showed that, for the primary outflow, a wide-angle flow has a relatively low velocity and a narrow outflow component has a relatively high velocity, as described in \S\ref{result-outflow}.

The previous observational studies imply that the primary outflow can be driven from the inner disk region because it has relatively high-velocity components and a well-collimated structure \citep[e.g.,][]{2019ApJ...871..221M}. In this case, the secondary outflow, which has a relatively low velocity, could be driven from the outer edge of the disk, where the disk is inclined relative to the inner disk, and the Keplerian velocity, roughly corresponding to the outflow velocity, is low. Since the velocity of the secondary outflow is slower than that of the primary outflow, the above scenario seems to explain the primary and secondary outflows. However, if this is the case, the outer disk region is inclined relative to the inner disk region by 90 degrees in the plane of the sky. 
Since the inclination angle between the inner and outer disks is significant, we are still determining whether such a disk is realized.
Considering the inclination of each outflow in three dimensions, the angle between the primary and secondary outflows can be less than 90 degrees. 
%In addition, recent MHD simulations have shown that the outflow propagation direction can be changed by $>90$ degrees when the rotation axis of a prestellar cloud core is misaligned with the global magnetic fields (e.g., see Figs.~10--12 of \citealt{machida2020a}).

\subsubsection{Outflows driven at different epochs}
Another possibility is that a protostar in Source\,A drives the primary and secondary outflows at different epochs. As described above, our $\mathrm{C^{18}O}$\,($J$ = 2--1) data (see Figure\,\ref{c18o_ch}) and previous observations \citep{Bergner2020} showed rotational motions in the east and west directions toward the peak position of the thermal dust emission. Thus, the rotation axis is expected to be parallel to the axis of the primary outflow at present. If this is the case, it is considered that a change of the disk normal direction caused a positional change of the primary and secondary outflows. The rotation axis of the disk was parallel to the secondary outflow in the past, and then it changed direction to the axis of the primary outflow. Thus, in this scenario, the primary outflow is currently being driven, while the secondary outflow was previously driven.

This scenario could be justified by recent simulation studies \citep{hirano2020,machida2020a} as described above. When the rotation axis of a prestellar cloud is not perfectly aligned with the global magnetic field, the disk normal direction changes with time. As a result, the outflow direction also changes with time (\citealt{machida2020a}). Magnetic braking transports the angular momentum of the infalling envelope. The efficiency of magnetic braking depends on many factors, such as the strength and configuration of the magnetic field and the disk and protostellar masses \citep{hirano2020}. In addition, the outflow and magnetic braking directly transport the disk angular momentum. Therefore, the disk normal direction changes with time, resulting in a directional change of the outflow \citep{hirano2019}. Thus, a single protostellar system can explain the secondary outflow without considering a binary system. To validate this scenario, we need to determine the configuration of the magnetic field using polarization observations.

As described in \S\ref{sec:intro}, a secondary outflow can be confirmed in other objects. IRAS 15398-3359 also shows a secondary outflow \citep{Okoda2021}. \citet{2018ApJ...864L..25O} estimated a protostellar mass of 0.007\,$M_{\odot}$ embedded in IRAS 15398-3359. They also suggested that the disk normal and the outflow direction can be easily changed in a very early mass accretion stage with episodic mass accretion. 
Indeed, \cite{Jorgensen2013ApJ} reported an accretion burst for IRAS 15398-3359 during the last 100--1000 years, possibly supporting the scenario described in \cite{Okoda2021}. We, therefore, estimated the central protostellar mass in Source\,A by assuming Kepler motion using $\mathrm{C^{18}O}$\,($J$ = 2--1) data to verify the possibility of episodic mass accretion and subsequent change of disk normal and outflow directions. We derived the central protostellar mass using the Keplerian velocity, $v_{\mathrm{kep}} = (GM_{*}/r)^{1/2}$, where $G$, $M_{*}$, and $r$ are the gravitational constant, central stellar mass, and radius from the central protostar, respectively. Our $\mathrm{C^{18}O}$\,($J$ = 2--1) image shows the emission extending to $\sim$ 400\,au with a maximum relative velocity of 2\,\kms. Thus, the central protostellar mass of Source\,A is estimated to be $\sim1.8\,M_{\odot}$. It is challenging for such a massive object to change the direction of the disk and outflow in a single protostellar system. Future observations resolving the roots of the outflows will determine the driving mechanism of these two outflows.

\vspace{5pt}
Considering all the evidence discussed in Sections\,\ref{sec:binary} and \ref{sec:single}, we concluded that Ser-emb\,15 is more likely to be a single system with a protostar driving both the primary and secondary outflows rather than a binary system. 
Our conclusion is supported by the absence of a confirmed protostar in the weak continuum intensity source (Source\,B), asymmetric continuum structure, and the agreement with simulation results on the outflow behavior around single protostars. Our study emphasizes the complexity and diversity in the early star formation stage. 
Finally, we stress the importance of high-resolution observations and detailed theoretical modeling to understand the driving mechanisms of outflows in such complicated systems.

\subsection{Non detection of a western lobe of the secondary outflow } \label{sec:non-detect}
\begin{figure}[ht]
  \centering
  \includegraphics[width=13cm]{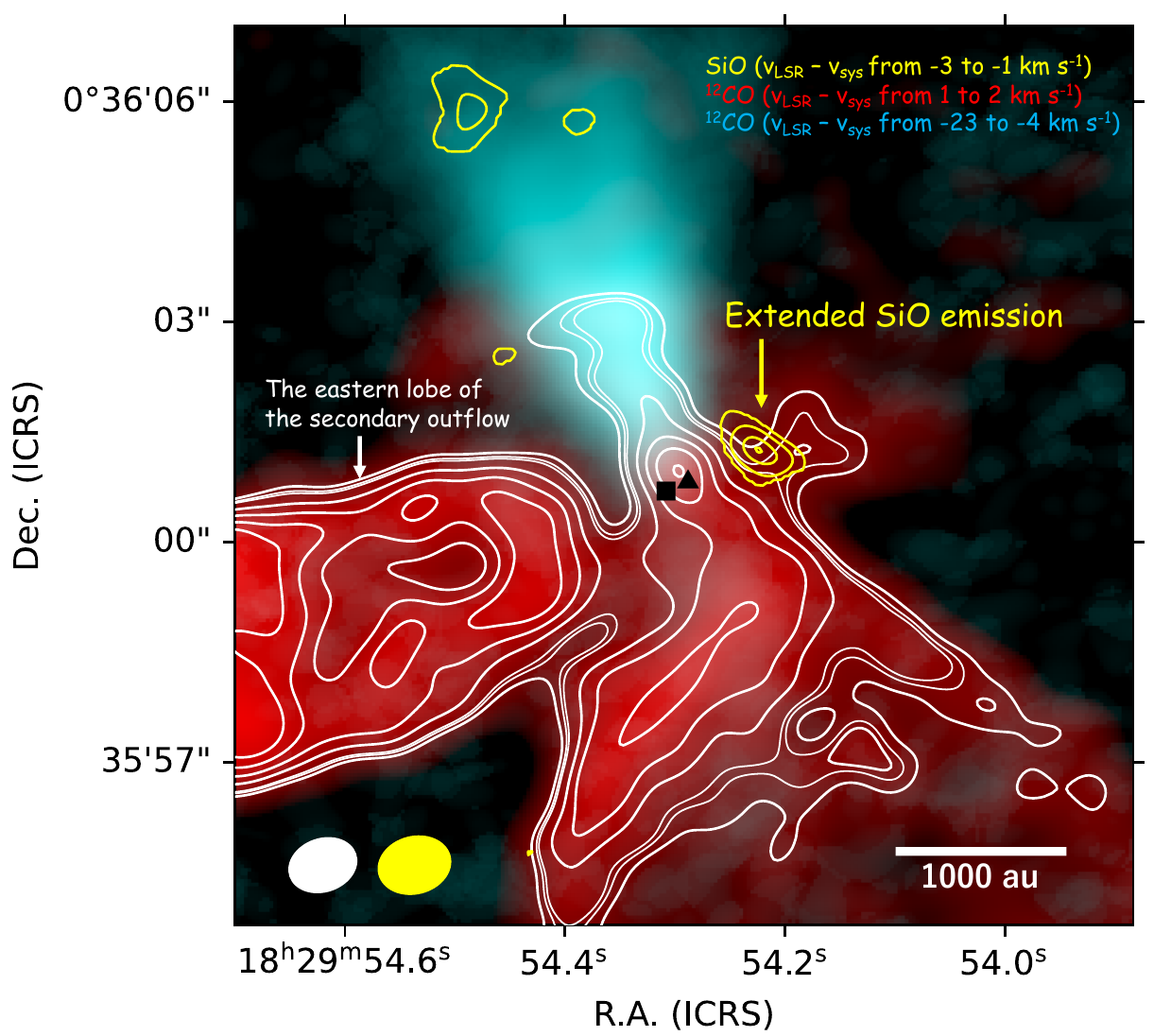}
  \caption{Comparison of spatial distributions between SiO\,($J$=5--4) and $^{12}$CO\,($J$=2--1) emission. The blue color represents the integrated intensity of the blue-shifted $^{12}$CO\,($J$=2--1) emission with the relative velocity in the range from $-$4 to $-$23\,\kms. The red color and white contours represent the integrated intensity of the red-shifted $^{12}$CO\,($J$=2--1) emission with the relative velocity in the range from 1 to 2\,\kms. The white contour levels are [12, 14, 15, 20, 25, 30, 35] $\times 1\sigma$ ($1\sigma = 6.5$\,\mjy). The yellow contours represent the integrated intensity of the blue-shifted SiO($J$=5--4) emission with the relative velocity in the range from $-$1 to $-$3\,\kms. The yellow contour levels are [5, 7, 9, 10] $\times 1\sigma$ ($1\sigma = 7.5$\,\mjy). The black square and triangle correspond to the peak positions of the dust continuum sources, Source\,A and B, respectively. The white and yellow ellipses at the bottom-left corner represent the synthesized beam size of the $^{12}$CO\,($J$=2--1) and SiO\,($J$=5--4) emission, respectively.
  }
  \label{fig:shock}
\end{figure}

As described in Section\,\ref{result-outflow}, we could not detect the blue-shifted CO lobe of the secondary outflow, which is expected to be located at the western side of Ser-emb\,15. 
Thus, the secondary outflow only has a one-side (or red-shifted) CO lobe. 
We discuss the non-detection of the blue-shifted lobe by comparing the CO and SiO emission presented in Figure\,\ref{fig:shock}. 
We can confirm detection of the blue-shifted SiO emission traced by the yellow contours in Figure\,\ref{fig:shock}, on which the blue-shifted (blue) and red-shifted (red) CO emission are superimposed. 
In the figure, the white contours represent the integrated intensity of the red-shifted CO emission and indicate that the red-shifted CO emission is associated with Ser-emb\,15.
In addition, the red-shifted CO emission also extends slightly to the western side of Ser-emb\,15. 
The CO emission has a relative velocity of $\sim2$\,\kms\ and it extends to 4\farcs0 ($\sim$1744\,au) from Ser-emb\,15. 
The extended SiO emission (yellow contours) is also located at the western side of Ser-emb\,15.
The peak position of the SiO emission is on the secondary outflow axis with P.A. = $+$110 degrees.

The SiO emission has a relative velocity in the range from $\sim -$1 to $-$3\,\kms, which is very close to the relative velocity of the red-shifted CO emission located on the same side.
By comparing the spatial and velocity distributions of the CO and SiO emission, we speculate that the extended SiO emission could trace a shocked region resulting from the collision between the gas expelled from Ser-emb\,15 and the surrounding dense material on the western side of Ser-emb\,15.
The surrounding dense material may be a circumstellar or circumbinary disk, discussed in \S\ref{sec:binary} and \ref{sec:single}.
Thus, the non-detection of the blue-shifted CO lobe can be attributed to the obstruction of the dense gas distributed on the western side. 
Recent MHD simulations with turbulence also showed that the morphology of the outflow lobes is asymmetric with respect to the disk plane and that the two outflow lobes are not the same (see Figure\,9 in \citealt{Matsumoto2017}). 

If the extended SiO emission located on the western side of Ser-emb\,15 is caused by the collision, the CO emission could also be detected with a similar velocity range as the SiO emission. 
However, the CO emission with the corresponding velocity was not detected even at the $1\sigma$ (= 6.5\,\mjy) emission level. 
There are some possibilities to explain the non-detection.
We propose two possibilities below. 
(1) Absorption of the CO emission by the surrounding dense gas in the line-of-sight direction, when the SiO component travels with the blue-shifted velocity and the material formed by the collision between the outflow and the surrounding gas is located in the front of the outflow lobe.
Such self-absorption obstruction generally occurs near the systemic velocity. However, \cite{Tokuda2016ApJ} reported an example where a redshifted HCO$^{+}$\,($J$ = 3--2) outflow was not detected in a more optically thick CO\,($J$ = 3--2) line in the MC27/L1521F dense core system in Taurus. This situation is similar to our case, where the outflow was detected in the SiO but not in the CO\,($J$ = 2--1). 
(2) Excitation of CO molecule due to the collision of the outflow with the surrounding dense envelope.
The collision, which could be detected in SiO emission, can heat and compress the surrounding gas, raising its kinetic temperature to over 1000\,K and its $\mathrm{H_2}$ gas number density to over $10^5\,\mathrm{cm^{-3}}$ \citep{Gusdorf2008A&A...482..809G, Godard2019A&A}.
In such a hot and dense environment, the CO\,($J$ = 2--1) can be excited, but the higher-$J$ CO line emission would be much stronger. Therefore, the CO\,($J$ = 2--1) emission might not be detected with the observational sensitivity in this study (6.5\,\mjy\,at the 1.0\,\kms\,resolution).
%In summary, with the current data, we speculate that the non-detection of the blue-shifted CO lobe of the secondary outflow is due to (1) the absorption by the surrounding gas in the line-of-sight direction, (2) the excitation of the CO\,($J$ = 2--1) state to the higher-$J$ state. Thus, t
To further reveal the characteristics of the secondary outflow, we need more data on the lines tracing the shocked gas, the higher-$J$ CO lines, and the optically thin lines.

\section{conclusion} \label{sec:conclusion}
We have presented the detection of the secondary outflow associated with a Class I source, Ser-emb\,15, using ALMA 1.3\,mm continuum data at an angular resolution of 0\farcs1 ($\sim$44\,au) and $\mathrm{^{12}CO}$\,($J$ = 2--1), $\mathrm{SiO}$\,($J$ = 5--4), and $\mathrm{C^{18}O}$\,($J$ = 2--1) line data at an averaged angular resolution of  0\farcs76 ($\sim$331\,au). Our results and discussions are detailed as follows:

\begin{enumerate}
    \item Our 1.3\,mm continuum image spatially resolved a previously identified continuum source, associated with Ser-emb\,15, into two compact sources and we named them Source\,A and B. Projected sizes of Sources\,A and B are measured to be 137\,au and 60\,au, respectively. Assuming a dust temperature of 20\,K, we estimate the dust mass to be $2.4 \times 10^{-3} \,M_{\odot}$ for Source\,A and $3.3 \times 10^{-4}\,M_{\odot}$ for Source\,B. From the CO data, we identified two pairs of molecular outflows consisting of three lobes, namely primary and secondary outflows. The secondary outflow was identified for the first time in this study. Its elongation is approximately perpendicular to the elongation direction of the primary outflow in the plane of the sky. The primary outflow axis is perpendicular to the major axis of the extended continuum structure within which Sources\,A and B are located. In addition, from the SiO data, we first identified high-velocity outflow associated with Ser-emb\,15, and its positional angle is consistent with that of the primary outflow. If ignoring the secondary outflow and the low-intensity peak (Source\,B), the Ser-emb\,15 system is a typical disk-outflow system confirmed in recent observations and numerical simulations. However, the presence of the secondary outflow suggests a distinctive star formation process in this system.

    \item To explain the secondary outflow, we have proposed two possible scenarios of (1) binary and (2) single-star systems.
    For the binary system scenario (1), the presence of a weak continuum source (Source\,B) supports  the possibility of a binary system. In this scenario, Source\,A alone could drive both outflows, with changes in the direction of the circumstellar disk resulting from interactions with its companion, Source\,B. The driving mechanism is supported by both observational and numerical simulation studies (e.g., \citealt{2013MNRAS.428.1321T, bally2015}). In addition, for the binary system scenario (1), the primary and secondary outflow could be driven by Source A and B, respectively, given the previously reported correlation between $L_{\mathrm{bol}}$ and outflow parameters \citep{Wu2004A&A...426..503W}. However, it is doubtful whether a binary companion is actually embedded in Source\,B due to its very weak intensity.
    
    In the single-star system scenario (2), both outflows can be driven by a protostar embedded in Source\,A at the same or different epochs if the observed asymmetry in the continuum emission is attributed to a warped disk surrounding Source\,A. Different outflow morphologies support the same-epoch driving scenario due to different outflow driving mechanisms in inner and outer disk regions proposed by recent observational and theoretical studies (e.g., \citealt{2022arXiv220913765T, 2019ApJ...871..221M}). The angle between the primary and secondary outflows is about  90 degrees in the plane of the sky, implying that the rotation axis of the outer disk region is perpendicular to that of the inner disk region. Although forming such a disk is challenging, their orientation could be less than 90 degrees in three dimensions. Another scenario, in which the primary and secondary outflows appear at different epochs in the single-star system, is supported by a directional change in the rotational axis of the disk, which has also been proposed by both observational and theoretical studies (e.g., \citealt{hirano2020, Okoda2021}). In this study, the central protostellar mass of Source\,A was estimated to be $\sim$1.8\,\msun, assuming Keplerian motion.  The protostellar mass embedded in Source\,A may be too massive to change the direction of the rotational axis in a single-star system. However, our data cannot accurately analyze rotational motion with the precision required to determine the mass.

\end{enumerate}

In this study, we could not conclusively determine the driving mechanism of the secondary outflow.
However, the single-star system scenario could better explain the origin of the mechanism than the binary scenario, supported by the absence of a confirmed protostar in the weak continuum intensity source (Source B), the presence of an asymmetric continuum structure, and the agreement with simulation results on the outflow behavior around a single protostar.
Since the secondary outflow is considered to be related to disk formation or binary formation, we need higher spatial-resolution observations to understand the disk or binary formation process better.

\section{acknowledgments}
We deeply acknowledge the referee for the careful reading and constructive comments that have helped to improve this manuscript. We thank R. Kawabe for helping us with analyzing the observational data used in this paper. This paper makes use of the following ALMA data: ADS/JAO.ALMA\#2019.1.01792.S. ALMA is a partnership of ESO (representing its member states), NSF (USA) and NINS (Japan), together with NRC (Canada), MOST and ASIAA (Taiwan), and KASI (Republic of Korea), in cooperation with the Republic of Chile. The Joint ALMA Observatory is operated by ESO, AUI/NRAO and NAOJ.
This work was supported by a NAOJ ALMA Scientific Research grant (No. 2022-22B), Grants-in-Aid for Scientific Research (KAKENHI) of the Japan Society for the Promotion of Science (JSPS; grant Nos. JP21K13962, JP21K03617, JP21H00049, and JP21H00046).
This work was also supported by the Kyushu University Leading Human Resources Development Fellowship Program.

\bibliography{ser-emb15}{}
\bibliographystyle{aasjournal}

\appendix
\section{$\mathrm{C^{18}O}$\,($J$ = 2--1)} \label{app_C18O}
\begin{figure}[ht]
 \begin{minipage}[b]{\linewidth}
 \centering
  \includegraphics[width=17cm]{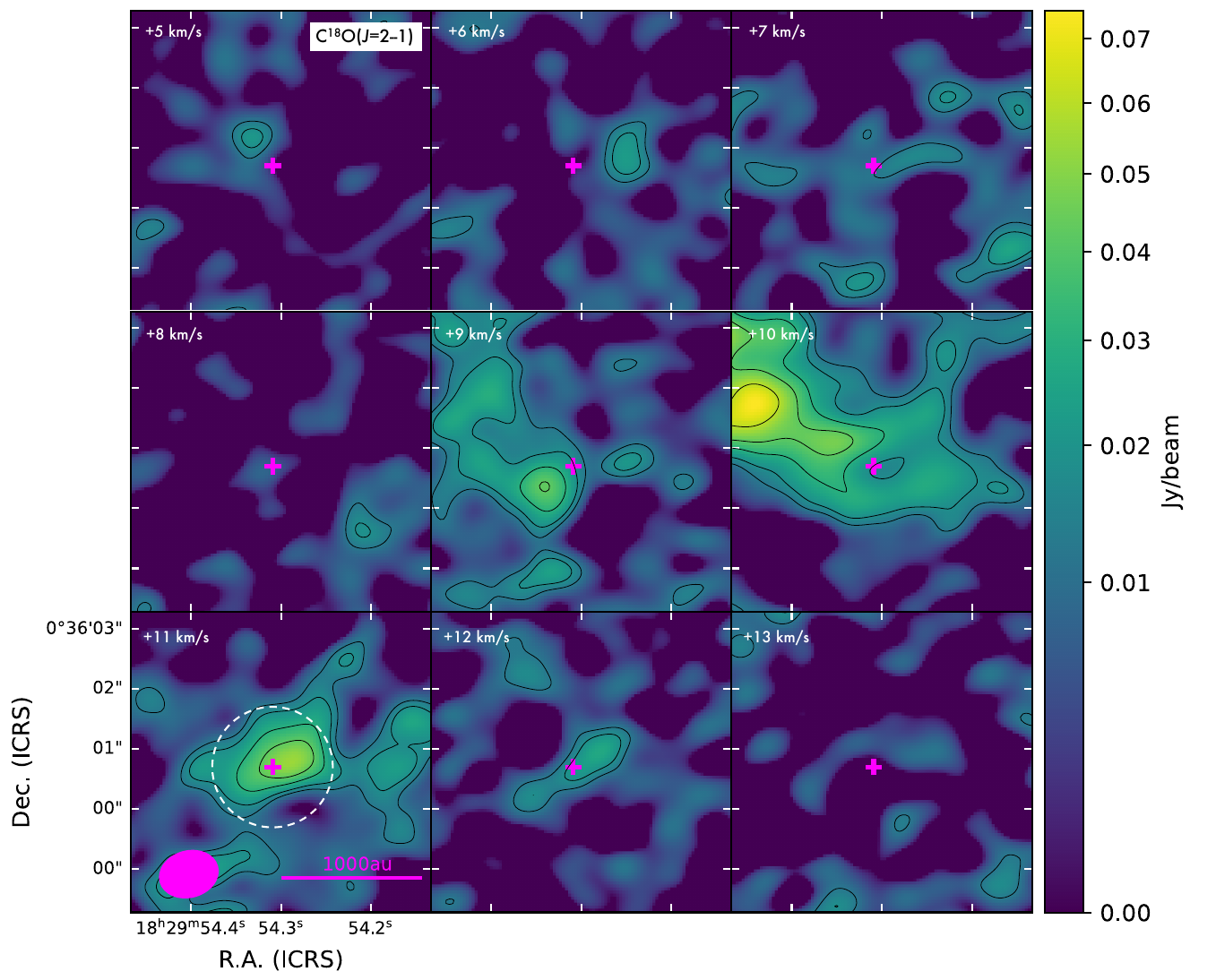}
  \end{minipage}\\
 \begin{minipage}[b]{\linewidth}
 \centering
  \includegraphics[width=15cm]{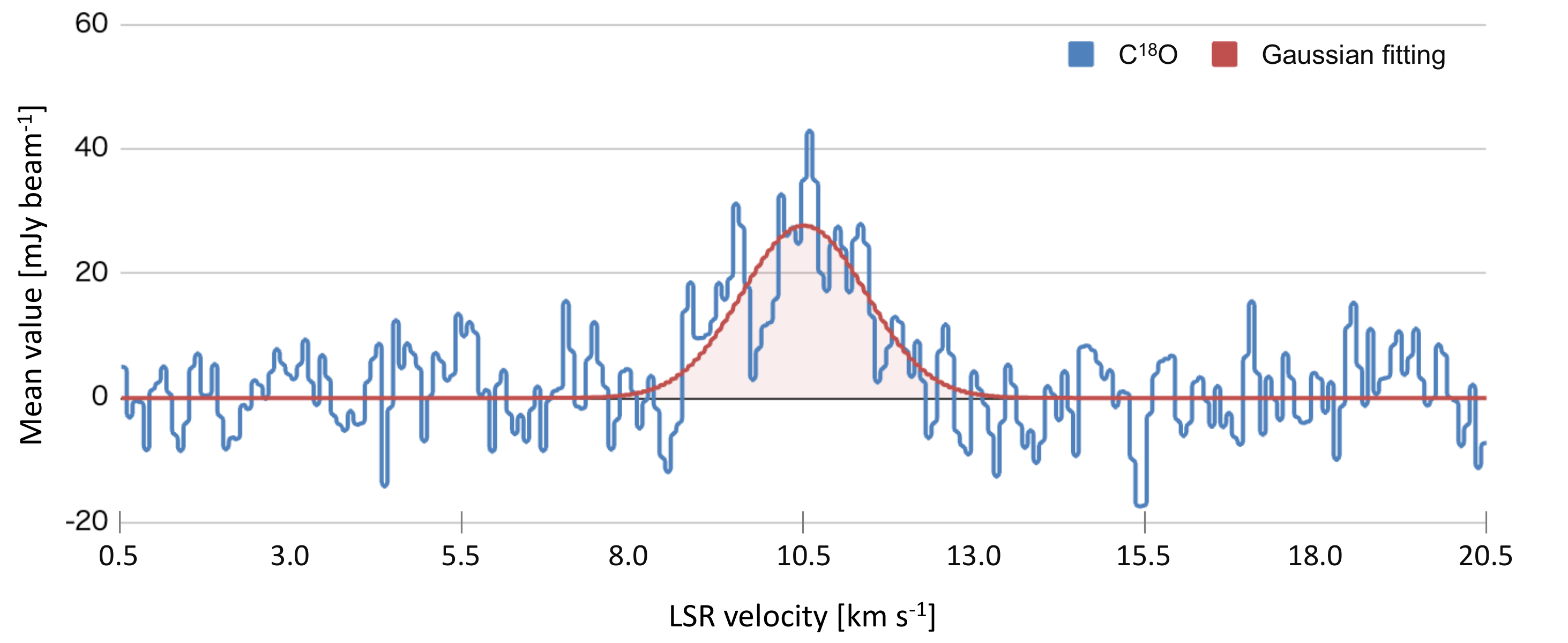}
 \end{minipage}
 \caption{\textbf{Top}: $\mathrm{C^{18}O}$\,($J$ = 2--1) channel map in the LSR velocity range from 5 to 13\,$\mathrm{km\,s^{-1}}$. The color scale and black contours represent the $\mathrm{C^{18}O}$ emission with a velocity resolution of 1.0\,$\mathrm{km\,s^{-1}}$. The magenta cross is the peak flux position for Source\,A in the 1.3\,mm continuum emission. The black contour levels are [2, 3, 5, 7, 10] $\times\,1\sigma$ ($1\sigma = 6.0$\,\mjy). The synthesized beam for the $\mathrm{C^{18}O}$ emission is denoted by the magenta ellipse in the panel at the bottom left corner.
 \textbf{Bottom}: $\mathrm{C^{18}O}$\,($J$ = 2--1) line spectrum obtained from a circular area with a radius of 0\farcs5 ($\sim$ 218\,au) centered on the peak flux position for Source\,A denoted by the dashed white line in the 11\,$\mathrm{km\,s^{-1}}$ channel of the top panel. The solid blue and red lines represent the line spectrum from the $\mathrm{C^{18}O}$ data with a velocity resolution of 0.1 \kms\, (rms = 30\,\mjy) and the results of Gaussian fitting, respectively. The Gaussian fitting shows that the LSR velocity at the peak intensity is 10.5 \kms.}
  \label{c18o_ch}
\end{figure}

We also present channel maps for $\mathrm{C^{18}O}$\,($J$ = 2--1) to show the rotational motions around Sources\,A and B in the top panel of Figure\,\ref{c18o_ch}.
The figure shows a weak emission extending 1\farcs8 ($\sim$ 800\,au) in the south-east and 1\farcs1 ($\sim$ 500\,au) in the north-west directions centered at Sources\,A and B in the LSR velocity range is from 9 to 12\,\kms at 3$\sigma$ ($1\sigma$ = 6.0\,\mjy).
The spatial resolution of the $\mathrm{C^{18}O}$ data is 0\farcs5 ($\sim$ 218\,au), which is not sufficient to spatially resolve Sources\,A and B. 
Thus, we cannot identify whether the emission originates from Source\,A and/or Source\,B. 
Figure\,\ref{c18o_ch} bottom shows a line spectrum of the $\mathrm{C^{18}O}$\,($J$ = 2--1) data. The Gaussian fitting shows that the central LSR velocity is 10.5\,($\pm$0.2)\,\kms. Hence, we used a systemic velocity of Ser-emb\,15 of 10.5 \kms.

\section{uncertainty of the dust mass} \label{app_dust}
To estimate the dust masses, in \S\ref{result-cont}, we assumed the optically thin dust emission and adopted the dust temperature of 20\,K, which seems to be a typical dust temperature around protostars \citep{Andrews2005ApJ}. 
However, there are uncertainties, such as the dust opacity and dust temperature, to estimate the dust mass. 
Thus, we discuss their influences on the dust mass in this section.

Firstly, although we used the dust opacity ($\kappa_{\mathrm{1.3mm}}$ = 0.899\,$\mathrm{cm^2\,g^{-1}}$) referring to \cite{ossenkopf1994}, \cite{Bergner2020} adopted $\kappa_{\mathrm{1.1mm}} = 2.62\,\mathrm{cm^2\,g^{-1}}$ following the power-law scaling of \cite{Beckwith1990AJ} with a power-law index $\beta$ = 1.
Using the same power-law scaling with $\beta$ = 1, we derived $\kappa_{\mathrm{1.3mm}} = 2.30\,\mathrm{cm^2\,g^{-1}}$. 
Adopting the dust opacity $\kappa_{\mathrm{1.3mm}}$ with $T_{\mathrm{dust}}$ = 20\,K, the dust mass estimated for Sources A and B are smaller than our estimate (see \S\ref{result-cont}) by a factor of 2.5 ($M_{\rm dust, A} = 9.2 \times 10^{-4}\,\mathrm{M_{\odot}}$ and $M_{\rm dust, B} = 1.3 \times 10^{-4}\,\mathrm{M_{\odot}}$).
Thus, the effect of the dust opacity is not significant to determine the dust mass. 

The dust grains grow and the opacity varies as the protostellar system evolves. 
An early observational study by \cite{Motte1998} adopted the dust opacity $\kappa_{\mathrm{1.3mm}}$ = 0.5\,$\mathrm{cm^2\,g^{-1}}$ for pre-stellar sources, 1.0\,$\mathrm{cm^2\,g^{-1}}$ for Class\,0/I sources, and 2.0\,$\mathrm{cm^2\,g^{-1}}$ for Class\,II sources. Even using these values for $\kappa_{\mathrm{1.3mm}}$ with a fixed $T_{\mathrm{dust}}$ of 20\,K, the uncertainty of the total dust mass is within a factor of 2.3 ($M_{\rm dust, A} \simeq 1.1 \times 10^{-3}$ to $4.3 \times 10^{-3}\,\mathrm{M_{\odot}}$ and $M_{\rm dust, B} \simeq 1.5 \times 10^{-4}$ to $5.9 \times 10^{-4} \,\mathrm{M_{\odot}}$).
Thus, the dust mass estimated in \S\ref{result-cont} may be a few times smaller or larger.

Finally, we discuss the optical thickness of the dust thermal emission. 
The dust emission from disks in submillimeter wavelength can be optically thick, resulting in underestimation of dust masses. 
We can estimate the dust temperature as 
\begin{equation}
  T_{\mathrm{dust}} = T_{\mathrm{0}}\left( \frac{L_{\mathrm{bol}}}{1L_{\odot}} \right)^{0.25},
  \label{eq2}
\end{equation}
where $T_{\mathrm{0}} = 43$\,K \citep{Tobin2020ApJ} and $L_{\mathrm{bol}}$ is the bolometric luminosity.
The dust temperature $T_{\mathrm{dust}}= 43$\,K is reasonable for $\sim1L_{\odot}$ at the disk radius of $\sim50$\,au \citep{Whitney2003ApJ, Tobin2020ApJ}.
Substituting $L_{\mathrm{bol}} = 0.17$\,\lsun\,of Ser-emb\,15 \citep{Enoch2011} to equation\,(\ref{eq2}), we derive $T_{\mathrm{dust}} \simeq 30$\,K.
In addition, considering the optical thickness, \cite{Tobin2020ApJ} estimated dust temperatures using a radiative transfer model to determine dust disk masses around Class\,0/I sources.
They derived a dust temperature distribution, in which $T_{\mathrm{dust}}\sim$50\,K is estimated at the disk radius of 50\,au (see Appendix\,B of \citealt{Tobin2020ApJ}). 
The projected physical size of Sources\,A and B are $\sim$130\,au and 60\,au, respectively.
Thus, we could assume the maximum dust temperature as 50\,K. 
Assuming a fixed opacity $\kappa_{\mathrm{1.3mm}}=0.899$\,$\mathrm{cm^2\,g^{-1}}$ and a temperature in the range $T_{\mathrm{dust}}=30-50$\,K, 
the dust mass is estimated to be smaller than that calculated in \S\,\ref{result-cont} within a factor of 3.0
($M_{\rm dust, A} \simeq 7.9 \times 10^{-4}$ to $1.4 \times 10^{-3} \,\mathrm{M_{\odot}}$ and $M_{\rm dust, B} \simeq 1.1 \times 10^{-4}$ to $2.0 \times 10^{-4} \,\mathrm{M_{\odot}}$).

In summary, the estimated dust masses have uncertainties of a factor of three at most compared to the masses estimated with $T_{\mathrm{dust}}= 50$\,K and $\kappa_{\mathrm{1.3mm}}=0.899$\,$\mathrm{cm^2\,g^{-1}}$.
However, the uncertainties do not qualitatively change our conclusions.

\section{CO line emission} \label{app_CO}
\begin{figure}[ht]
  \centering
  \includegraphics[width=\textwidth]{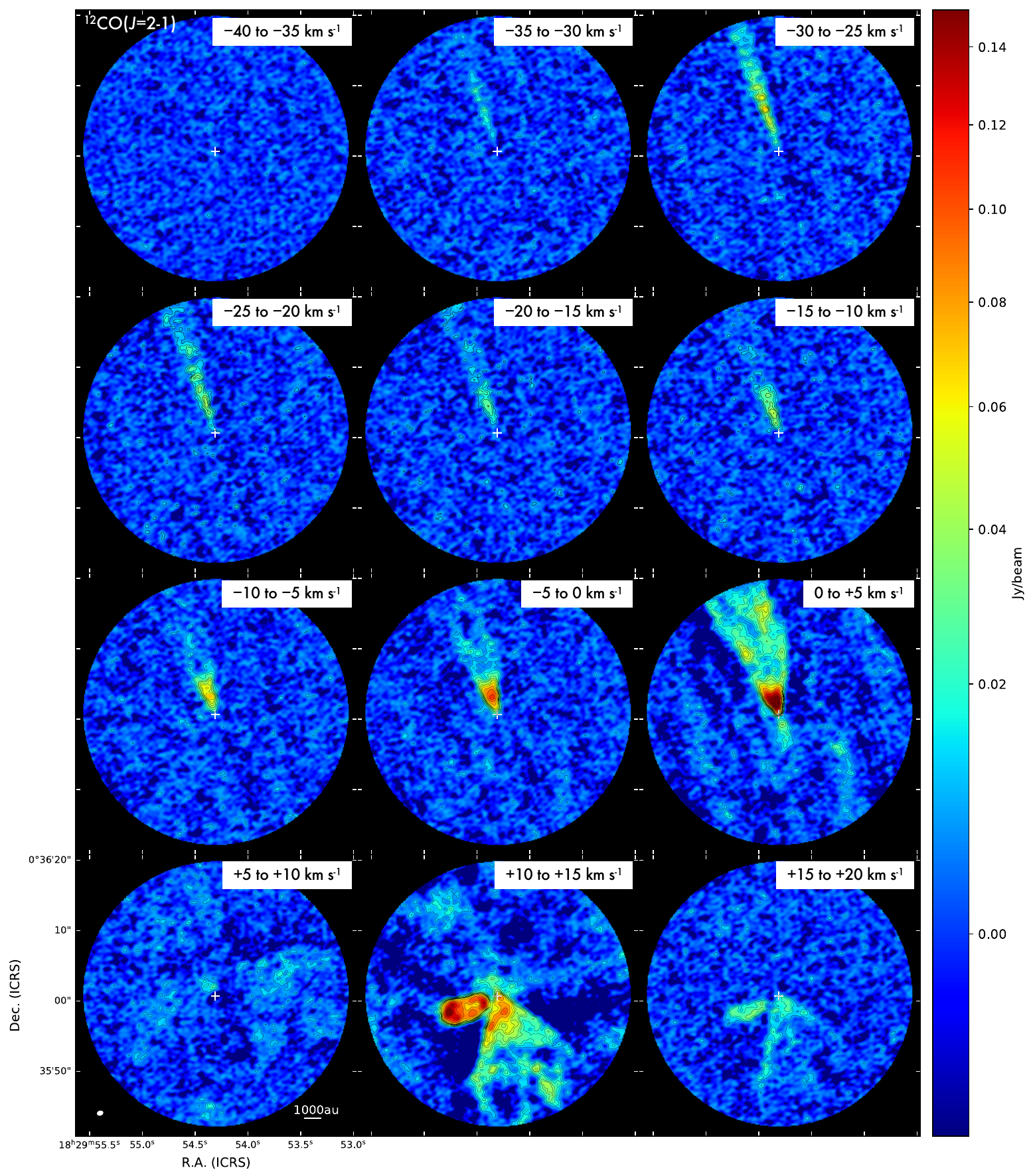}
  \caption{$^{12}$CO\,($J$=2--1) channel map in the LSR velocity range from $-$40 to 20\,$\mathrm{km\,s^{-1}}$. The color scale and black contours represent the CO emission. Each panel is obtained by combining the frequency channels to be in a velocity range of 5.0\,$\mathrm{km\,s^{-1}}$\,using the CASA task of \texttt{imrebin}. The white cross is the peak flux position for Source\,A in the 1.3\,mm continuum emission. The white contour levels are [3, 5, 7, 10, 12, 15, 17, 20, 25, 30, 35, 40, 45, 50, 55, 60]$\times\,1\sigma$ ($1\sigma = 3.0$\,\mjy). The synthesized beam for the CO emission is denoted by the white ellipse in the panel at the bottom left corner. The color bar on the right side is fixed in the range from $-$5.0$\times 10^{-3}$ to 0.15\,\jy.}
  \label{12co_ch_map_1}
\end{figure}

\begin{figure}[ht]
  \centering
  \includegraphics[width=\textwidth]{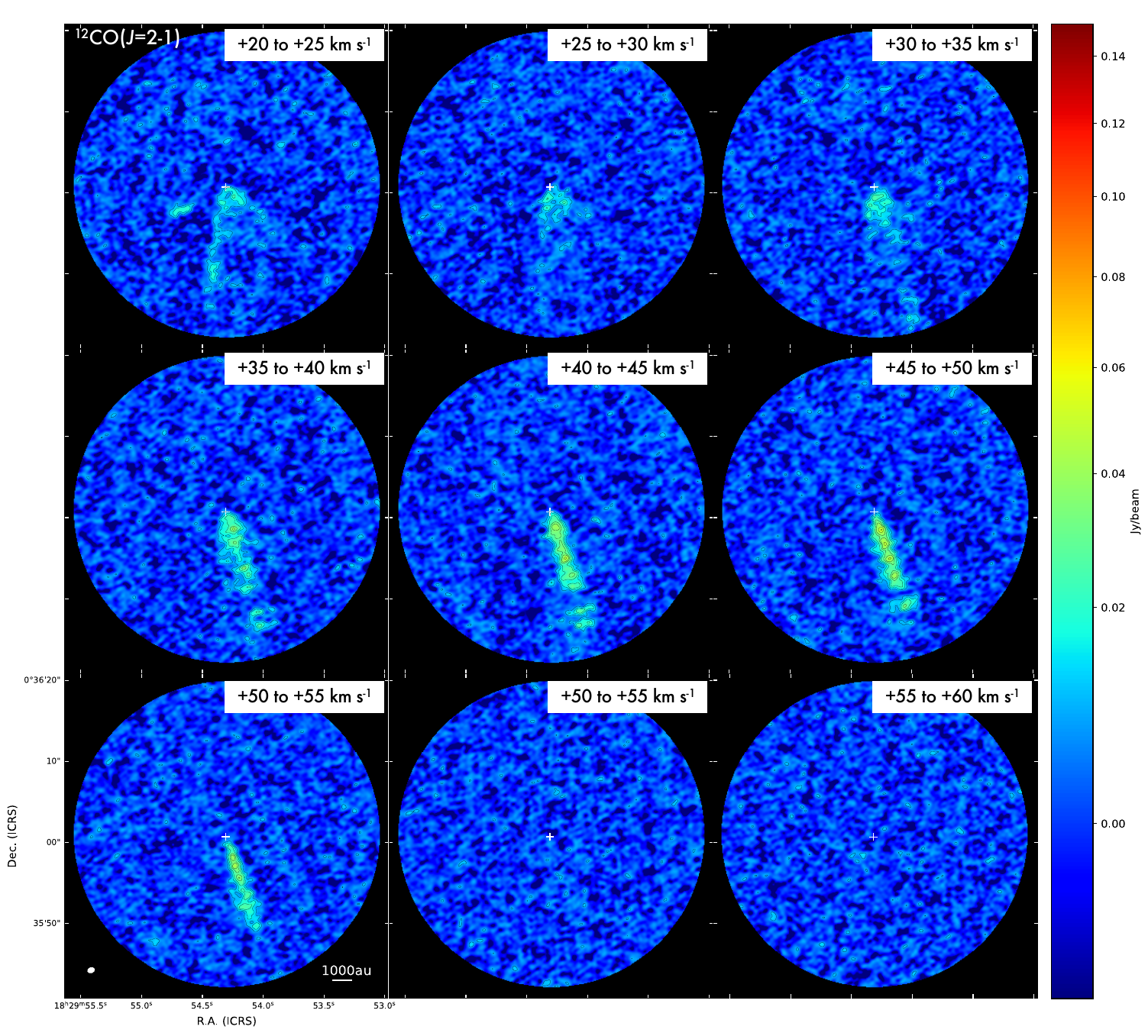}
  \caption{$^{12}$CO\,($J$=2--1) channel map in the LSR velocity range from 20 to 60\,$\mathrm{km\,s^{-1}}$. The color scale and black contours represent the CO emission. Each panel is obtained by combining the frequency channels to be in a velocity range of 5.0\,$\mathrm{km\,s^{-1}}$\,using the CASA task of \texttt{imrebin}. The white cross is the peak flux position for Source\,A in the 1.3\,mm continuum emission. The white contour levels are [3, 5, 7, 10, 12, 15, 17]$\times\,1\sigma$ ($1\sigma = 3.0$\,\mjy). The synthesized beam for the CO emission is denoted by the white ellipse in the panel at the bottom left corner. The color bar on the right side is the same as that in Figure~\ref{12co_ch_map_1}.}
  \label{12co_ch_map_2}
\end{figure}

To investigate the detailed spatial distributions, we present channel maps for the $^{12}$CO\,($J$ = 2--1) emission in Figures~\ref{12co_ch_map_1} and \ref{12co_ch_map_2}.

%\restartappendixnumbering
%\restartappendixnumbering

\section{SiO line emission} \label{app_SiO}
\begin{figure}[ht]
  \centering
  \includegraphics[width=\textwidth]{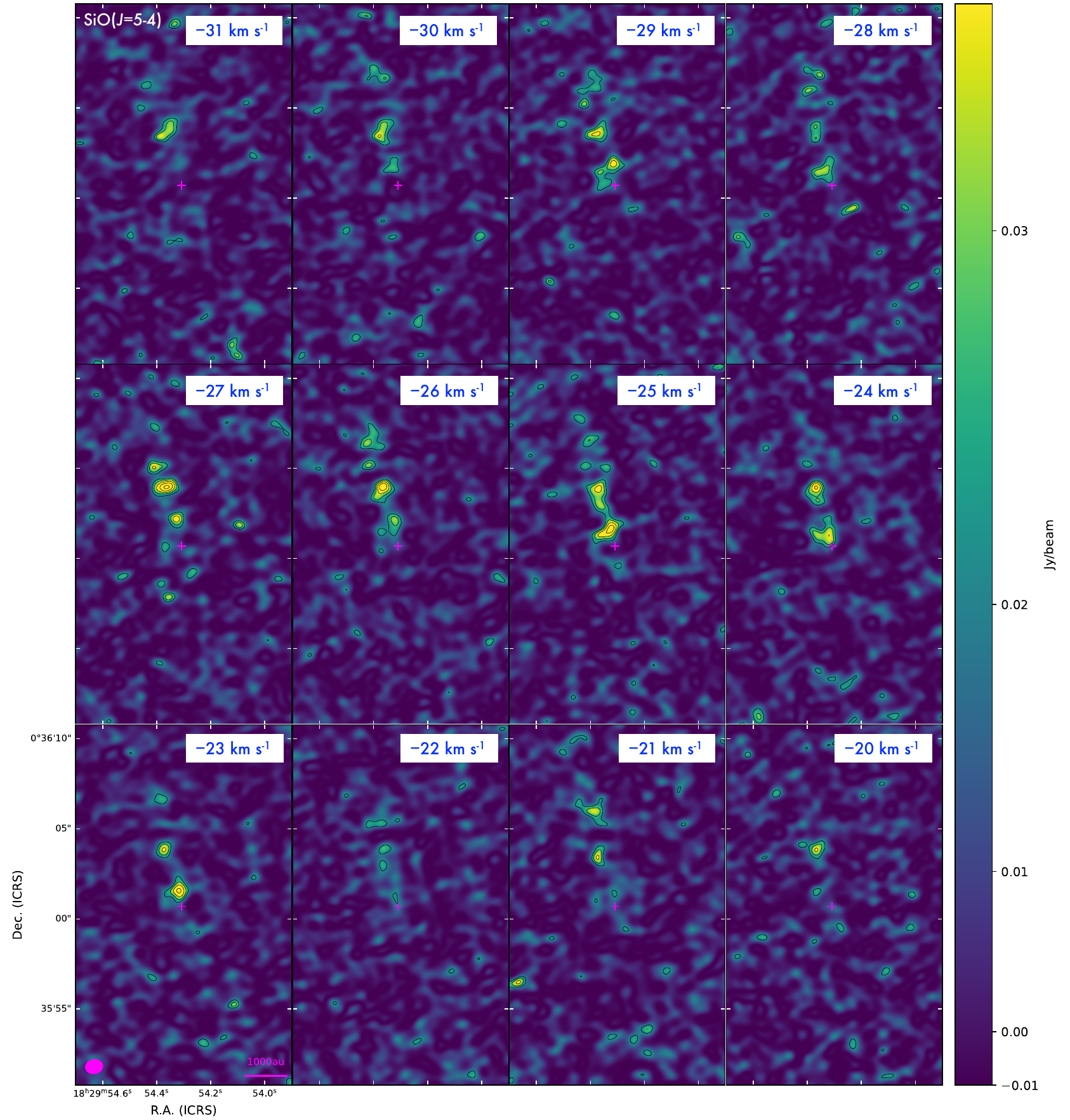}
  \caption{SiO\,($J$ = 5--4) channel map in the LSR velocity range from $-$31 to $-$20\,$\mathrm{km\,s^{-1}}$. The color scale and black contours represent the SiO emission with a velocity resolution of 1.0\,$\mathrm{km\,s^{-1}}$. The magenta cross is the peak flux position for Source\,A in the 1.3\,mm continuum emission. The black contour levels are [3, 4, 5, 6, 7] $\times\,1\sigma$ ($1\sigma = 7.0$\,\mjy). The synthesized beam for the SiO emission is denoted by the magenta ellipse in the panel at the bottom left corner.
  }
  \label{sio_ch_blue}
\end{figure}

\begin{figure}[ht]
  \centering
  \includegraphics[width=\textwidth]{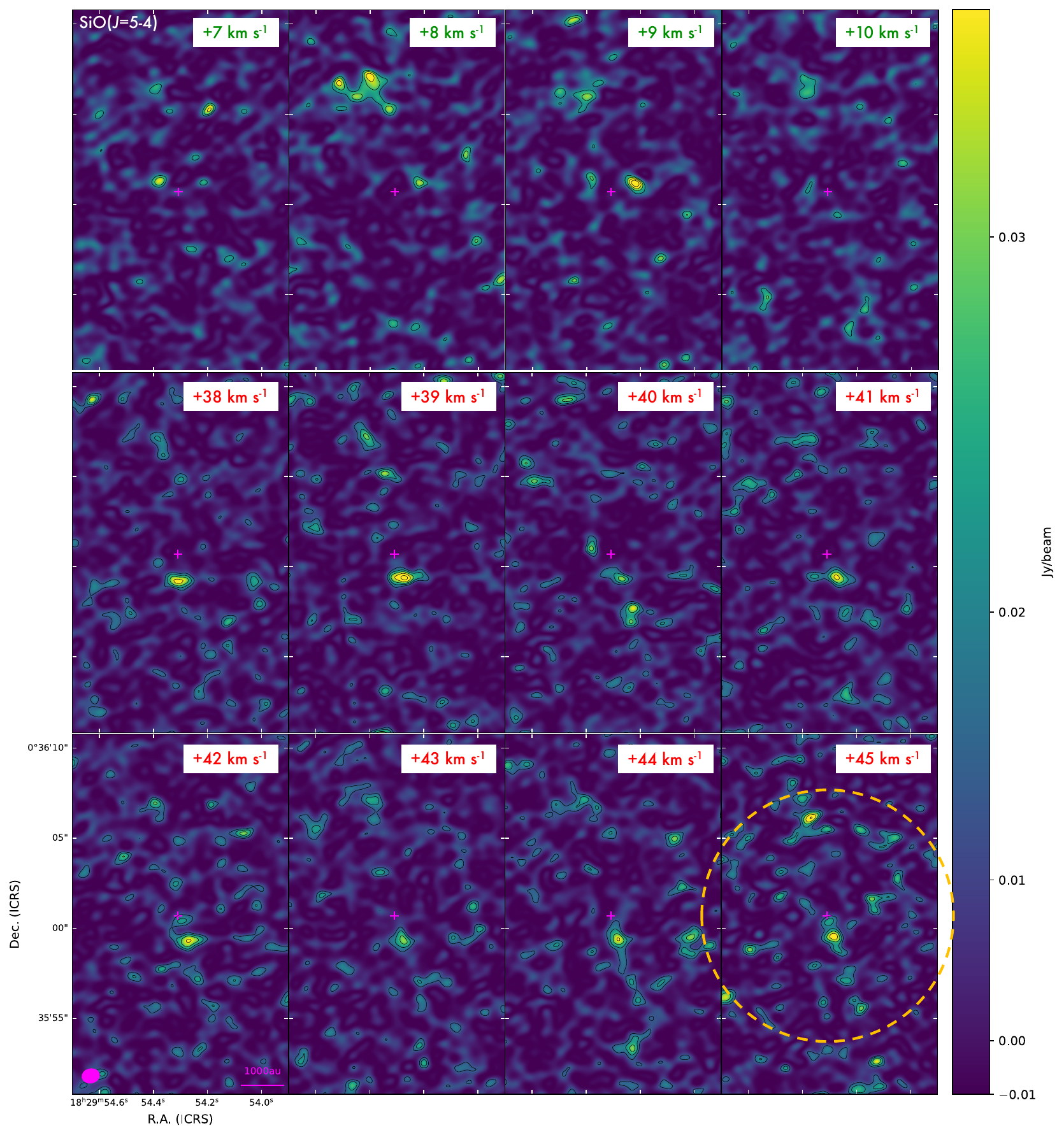}
  \caption{SiO\,($J$ = 5--4) channel map in the LSR velocity range from 7 to 10 and from 38 to 45\,$\mathrm{km\,s^{-1}}$. The color scale and black contours represent the SiO emission with a velocity resolution of 1.0\,$\mathrm{km\,s^{-1}}$. The magenta cross is the peak flux position for Source\,A in the 1.3\,mm continuum emission. The black contour levels are [3, 4, 5, 6]$\times\,1\sigma$ at the channels from 7 to 10\,$\mathrm{km\,s^{-1}}$ and [2, 3, 4, 5, 6]$\times\,1\sigma$ at the channels from 38 to 45\,$\mathrm{km\,s^{-1}}$ ($1\sigma = 7.0$\,\mjy). The dashed orange circle in the panel at the bottom right corner is the area where the line spectrum was measured (see Fig.\,\ref{sio_line_spectrum}). The radius of the blue circle centered on the magenta cross is 6\farcs9 ($\sim$3000\,au). The synthesized beam for the SiO emission is denoted by the magenta ellipse in the panel at the bottom left corner. The color bar on the right side is the same as in Figure~\ref{sio_ch_blue}.
  }
  \label{sio_ch_vsys_red}
\end{figure}

\begin{figure}[ht]
  \centering
  \includegraphics[width=\textwidth]{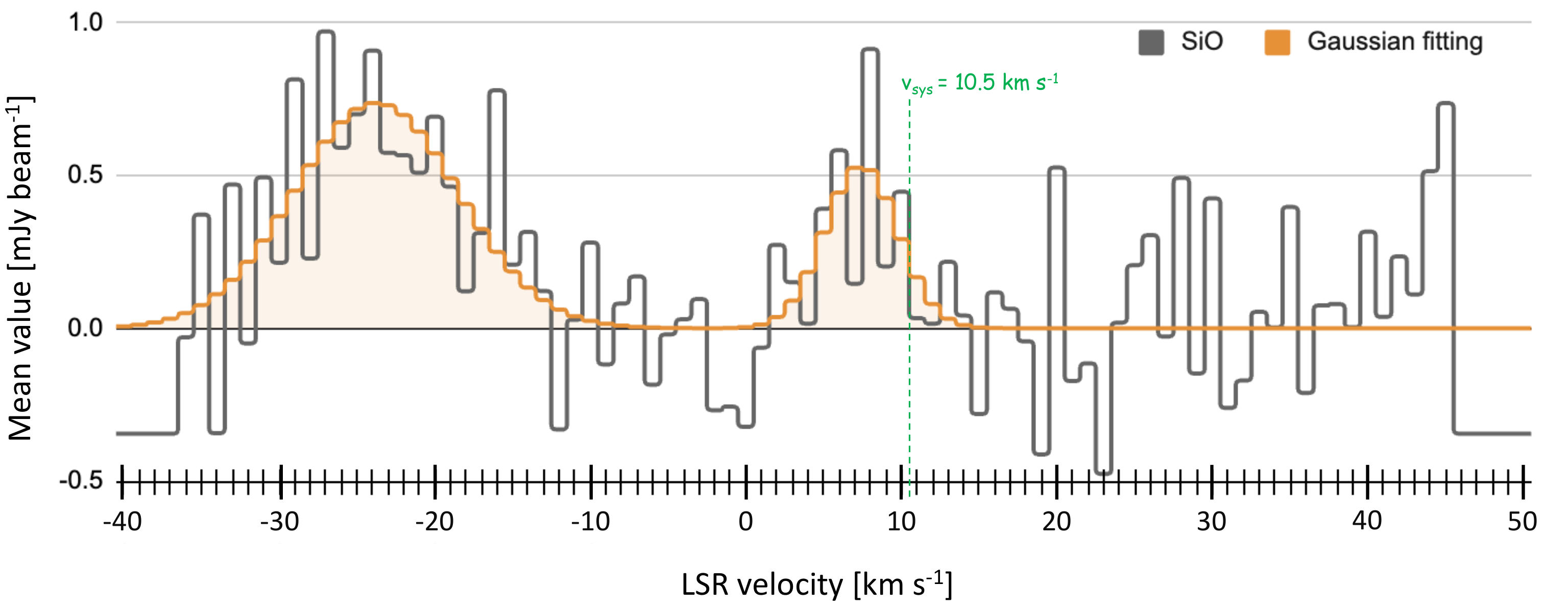}
  \caption{SiO\,($J$ = 5--4) line spectrum obtained from the integrated emission over the circular area denoted by the dashed blue line in the 45\,\kms\,channel of Figure\,\ref{sio_ch_vsys_red}. The solid gray line represents the line spectrum from the SiO data with a velocity resolution of 1.0\,\kms. The solid orange line is the result of Gaussian fitting with multiple components. The dotted green line corresponds to the systemic velocity of 10.5\,\kms. This SiO data contains the frequency channels with the LSR velocity from $-$36 to 45\,\kms.}
  \label{sio_line_spectrum}
\end{figure}

To investigate the detailed spatial distributions, we also present channel maps for the SiO\,($J$ = 5--4) emission in Figures~\ref{sio_ch_blue} and \ref{sio_ch_vsys_red}.
In addition, we present a line spectrum of the SiO data in Figure\,\ref{sio_line_spectrum} to clearly show the three velocity components (red-shifted velocity, blue-shifted velocity, and velocity close to the systemic velocity) classified in \S\ref{result-outflow} and denoted by different color contours in Figure~\ref{fig:outflow}. 
A multiple component Gaussian fitting shows two intensity bumps denoted by the orange line in Figure~\ref{sio_line_spectrum}. 
The LSR velocities at the peak intensity of these two bumps are $-$23.76 ($\pm$0.95)\,\kms\, and 7.42 ($\pm$0.88)\,\kms, respectively.
Note that we did not fit an intensity bump showing the red-shifted SiO component where the peak intensity is distributed at $\sim$45\,\kms, because there is no data higher than the LSR velocity of 45\,\kms\, as seen in Figure~\ref{sio_line_spectrum}.
The red-shifted velocity bump can be confirmed in the spectrum without the fitting, and the red-shifted component can also be identified in the SiO channel map (i.e., Figure\,\ref{sio_ch_vsys_red}).

\end{document}